\documentclass[reprint,bibnotes,amsmath,amssymb,prd,nofootinbib]{revtex4-1}

\usepackage{lipsum}
\usepackage{geometry}        
\usepackage{graphicx}
\usepackage{amssymb}
\usepackage{epstopdf,hyperref}
\usepackage{xcolor}
\usepackage{float}
\def\qrr@split@result#1 #2\@qrr@split@result{\edef\erfInput{#1}\edef\erfResult{#2}}
\newcommand*{\gnuplotErf}[2][\jobname.eval]{%
    \immediate\write18{gnuplot -e "set print '#1'; print #2, erf(#2);"}%
    \everyeof{\noexpand}
    \edef\qrr@temp{\input{#1} }%
    \expandafter\qrr@split@result\qrr@temp\@qrr@split@result
}

\newcommand{\beq}{\begin{equation}}
\newcommand{\eeq}{\end{equation}}
\newcommand{\bea}{\begin{eqnarray}}
\newcommand{\eea}{\end{eqnarray}}
\newcommand{\beann}{\begin{eqnarray*}}
\newcommand{\eeann}{\end{eqnarray*}}

\begin{document}

\title{On the Outskirts of Dark Matter Haloes}
\author{Alice Y. Chen}
\email{ay7chen@uwaterloo.ca}
\affiliation{Department of Physics and Astronomy, University of Waterloo, 200 University Ave W, N2L 3G1, Waterloo, Canada}
\affiliation{Waterloo Centre for Astrophysics, University of Waterloo, Waterloo, ON, N2L 3G1, Canada}
\affiliation{Perimeter Institute For Theoretical Physics, 31 Caroline St N, Waterloo, Canada}
 \author{Niayesh Afshordi}
 \email{nafshordi@pitp.ca}
\affiliation{Department of Physics and Astronomy, University of Waterloo, 200 University Ave W, N2L 3G1, Waterloo, Canada}
\affiliation{Waterloo Centre for Astrophysics, University of Waterloo, Waterloo, ON, N2L 3G1, Canada}
\affiliation{Perimeter Institute For Theoretical Physics, 31 Caroline St N, Waterloo, Canada}

\begin{abstract}
Halo models of large scale structure provide powerful and indispensable tools for phenomenological understanding of the clustering of matter in the Universe. While the halo model builds structures out of the superposition of haloes, defining halo profiles in their outskirts - beyond their virial radii - becomes increasingly ambiguous, as one cannot assign matter to individual haloes in a clear way. In this paper, we address this issue by finding a systematic definition of mean halo profile that can be extended to large distances - beyond the virial radius of the halo - and matched to simulation results. These halo profiles are compensated and are the key ingredients for the computation of cosmological correlation functions in an Amended Halo Model. The latter, introduced in our earlier work \cite{Chen:2019wik}, provides a more physically accurate phenomenological description of nonlinear structure formation, which respects conservation laws on large scales.  Here, we show that this model can be extended from the matter auto-power spectrum to the halo-matter cross-power spectra by using data from N-body simulations.  Furthermore, we find that this (dimensionless) definition of the compensated halo profile, $r^3 \times \rho(r)/M_{200c}$, has a near-universal maximum in the small range of $0.03-0.04$ around the virial radius, $r \simeq r_{\rm 200c}$, nearly independent of the halo mass.  The profiles cross zero into negative values in the halo outskirts - beyond 2-3$\times r_{\rm 200c}$ - consistent with our previous results. We provide a preliminary fitting function for the compensated halo profiles (extensions of Navarro-Frenk-White profiles), which can be used to compute more physical observables in large scale structure. 
\end{abstract}
\maketitle

\section{Introduction}
\label{sec:intro}
Probing properties of the large scale structure of the universe is an active area of research in cosmology.  While the interior structure of dark matter haloes can be well modelled with the Navarro-Frenk-White (NFW) or Einasto profiles \cite{Navarro:1995iw, Ludlow:2016qow}, the large scale distribution of dark matter in the outskirts of individual haloes is poorly understood in semi-analytic frameworks.  Previous studies have attempted to produce a model using effective field theory (EFT) and perturbation theory \cite{Schmidt:2015gwz, Seljak:2015rea, Hand:2017ilm, Philcox:2020rpe}, but non-linearities in structure formation make it difficult to extrapolate the models beyond $k\geq 1.0 {\rm Mpc}^{-1}$. For smaller scales, the Standard Halo Model (SHM) is often used as a phenomenological framework.  However, SHM suffers from pathologies that stem from not enforcing the conservation laws \cite{Cooray:2002dia}. One proposal to address these pathologies, namely the large-scale shot noise, has been to impose an exclusion radius for haloes \cite{2017PhRvD..96h3528G,Umeh:2021xqm}, but it is hard to see how this would distinguish between the conserved and non-conserved quantities, as the latter are expected to display shot noise on large scales. 

In order to address this issue in a systematic manner, we introduced an amended halo model (AHM) \cite{Chen:2019wik} with compensated halo profiles and fitted this to cold dark matter simulations by Takahashi et al. \cite{Takahashi:2012em} to model non-linear dark matter density power spectrum on scales of $10^{-2}~ {\rm Mpc}^{-1} \lesssim  k \lesssim 100~ {\rm Mpc}^{-1}$.  However, this analysis did not consider the profile's potential dependence on halo mass, nor did it include non-linear biasing in halo-halo correlations \cite{Mead:2020qdk}.  Consequently, the present study aims to address these deficiencies.  To do this, we adopt simulation data from the DarkEmu cosmological emulator suite \cite{Nishimichi:2018etk} - which uses Planck 2015 cosmology \cite{Ade:2015zua} - to develop a novel and systematic method to directly measure mean compensated profiles from simulated (or emulated) halo-matter and halo-halo correlations.  We find that the dimensionless compensated halo profiles all peak around the virial radius at a near-universal maximum even across different mass bins, which is quite a striking result.  We also find that it is possible to ``extrapolate'' the NFW profile beyond the virial radius by using two extra parameters to fit our compensated halo profile, and provide an approximate functional form of the profile (although more parameters are likely necessary if we want a highly accurate numerical fit).  Having such a fit for a semi-analytic framework allows us to make predictions for halo power on a larger scale (larger r and k range), beyond the resolution of current N-body simulations. 

While the physical meaning of the fitting parameters used here is yet to be determined, our main goal here is to show that a simple fit for the compensated halo profile does exist and can be used to predict the matter-halo cross-correlation, while avoiding the pathologies of the standard halo model.  Our fit matches NFW in the halo regions $r<r_{\rm virial}$, but is compensated in the outer regions $r>r_{\rm virial}$ \footnote{In this paper, the virial radius $r_{\rm virial}$ is taken to be the radius where density is 200 times the critical density of the universe, or equivalent to $r_{\rm 200c}$. Unless noted otherwise, the mass of a halo is also defined as the total mass contained within this radius.}.  This paper is then structured as follows: Section \ref{sec:AHM} outlines the amended halo model \cite{Chen:2019wik} for halo-matter cross correlations, Section \ref{sec:results} discusses our findings from the results of the simulation data from DarkEmu \cite{Nishimichi:2018etk}, and Section \ref{sec:conclusion} summarizes the results and potential avenues for future research.  The exact form of the fitting function we used, and its derivation and implications are outlined in the Appendix.

\section{Standard vs Amended Halo Models}\label{sec:AHM}

\indent In the standard halo model (SHM), the matter overdensity in the universe, $\delta_m({\bf x})$, is described as a superposition of individual halo profiles, that we refer to here as $u^j$: 
\beq
\delta_m({\bf x}) = \bar{\rho}^{-1} \sum_j M_j u_{\rm SHM}^j({\bf x - x}_j), \label{delta_r}
\eeq
in real space, and 
\beq
\delta_{m,{\bf k}} = \bar{\rho}^{-1} \sum_j M_j u_{{\rm SHM},\bf k}^j \exp(i{\bf k\cdot x}_j),
\label{delta_f}
\eeq
in Fourier space.  $M_j$ and ${\bf x}_j$ are the mass and position of the $j$-th halo, respectively, while $\bar{\rho}$ denotes the mean density of the universe.   

To get a complete understanding of matter distribution, we need to know both how matter is distributed within the individual halos, $u^j$'s, and how these halos are distributed throughout space. Furthermore, conservation laws, such as those of mass and linear momentum, require a fine balance between these two distributions, which are often hard to enforce in the SHM formulation (but see \cite{Schmidt:2015gwz}).  

In order to address this, we introduced the Amended Halo Model (AHM), where we split the nonlinear overdensities between the linear $\delta_L({\bf x})$ and halo contributions, revising the SHM equations (\ref{delta_r}) and (\ref{delta_f}) to be:
\beq
\delta_m({\bf x}) = \delta_L({\bf x})+ \bar{\rho}^{-1} \sum_j M_j u_{\rm AHM} ^j({\bf x - x}_j)
\label{delta_x}
\eeq
and
\beq
\delta_{m,{\bf k}} = \delta_{L,{\bf k}}+ \bar{\rho}^{-1} \sum_{j} M_{j}  u_{\rm AHM, \bf {k}}^{j} \exp(i{\bf k\cdot x}_j)
\label{delta_k}
\eeq
in real and Fourier spaces, respectively.  The nonlinear contribution to the halo profile, which is what we use to construct $u_{AHM}^j$, is what we will be calculating from simulations, since the contribution from the linear term is already included in $\delta_L (\bf{x})$.

Now, let us consider the overdensity of {\rm haloes} within a mass-bin $b$:
\beq
\delta^{b}_{\rm{halo}, \bf k} =  \frac{1}{\bar{n}^b_{\rm halo}}\sum_j {\cal N}^j_{b} \exp(i{\bf k\cdot x}_j), 
\label{delta_khalo}
\eeq
where ${\cal N}^j_{b}=1$ if the $j$-th halo is within the mass-bin $b$, but vanishes otherwise. Furthermore, $\bar{n}^b_{\rm halo}$ is the number density of haloes within the mass bin. 

The auto matter, auto halo, and halo-matter cross spectra can now be defined as:
\bea
P_{mm}(k) \equiv \frac{\langle \delta_{m, \bf k} \delta^*_{m,{\bf k}}\rangle  }{V}
\\
P^{bc}_{hh}(k) \equiv \frac{\langle \delta^{b}_{\rm{halo}, \bf k} \delta^{c*}_{\rm{halo},{\bf k}}\rangle  }{V}
\\
P^{b}_{hm}(k) \equiv \frac{\langle \delta^{b}_{\rm{halo}, \bf k} \delta^{*}_{m,{\bf k}}\rangle  }{V},
\eea
respectively, where $V$ is the volume of the simulation, and $b$ and $c$ stand for different halo mass bins.
Now, by multiplying equations (\ref{delta_khalo}) and (\ref{delta_k}), we find the cross-power spectra:

\bea
 P^b_{hm}(k) &&= b(\bar{M}_b) P_L(k)+ \frac{1}{\bar{n}^b_{\rm halo}\bar{\rho} V}\sum_{j} M_j  {\cal N}^j_{b} u_{\rm AHM, \bf {k}}^{j}  \nonumber\\
&&  \textcolor{purple}{+} \frac{1}{\bar{n}^b_{\rm halo}\bar{\rho} V}\sum_{j\neq l} M_j  {\cal N}^l_{b} u_{\rm AHM, \bf {k}}^{j} \exp[i{\bf k}\cdot ({\bf x}_j- {\bf x}_l)],\nonumber\\
\label{p_hm}
\eea

which we can write in the matrix form:
\begin{widetext}
\begin{eqnarray}
    \begin{pmatrix} P^1_{hm}(k) \\ ... \\ ... \\ P^q_{hm}(k)
\end{pmatrix} = P_L(k)\begin{pmatrix} b(\bar{M}_{1}) \\ ... \\ ... \\ b(\bar{M}_{q})\end{pmatrix} +\begin{pmatrix} 
\bar{M}_{1}/\bar{\rho} & 0 & ... & 0 \\  
0 &  \bar{M}_{2}/\bar{\rho} & ... & 0 \\
... & ... & ... & ... \\
0 & ... & ... & \bar{M}_{q}/\bar{\rho}
\end{pmatrix} \begin{pmatrix} u^1_{\rm AHM}(k) \\ ... \\ ... \\ u^q_{\rm AHM}(k)
\end{pmatrix} + 
\nonumber\\
 \begin{pmatrix} 
P_{hh}(k|\bar{M}_{1},\bar{M}_{1})& P_{hh}(k|\bar{M}_{1},\bar{M}_{2})& ... & P_{hh}(k|\bar{M}_{1},\bar{M}_{q}) \\  
P_{hh}(k|\bar{M}_{2},\bar{M}_{1})& P_{hh}(k|\bar{M}_{2},\bar{M}_{2})& ... & P_{hh}(k|\bar{M}_{2},\bar{M}_{q}) \\
... & ... & ... & ... \\
P_{hh}(k|\bar{M}_{q},\bar{M}_{1})& P_{hh}(k|\bar{M}_{q},\bar{M}_{2})& ... & P_{hh}(k|\bar{M}_{q},\bar{M}_{q})
\end{pmatrix} \times \begin{pmatrix} 
\mu_1 & 0 & ... & 0 \\  
0 &  \mu_2 & ... & 0 \\
... & ... & ... & ... \\
0 & ... & ... & \mu_q
\end{pmatrix} \begin{pmatrix} u^1_{\rm AHM}(k) \\ ... \\ ... \\ u^q_{\rm AHM}(k)
\end{pmatrix}.
\label{p_matrix}
\end{eqnarray}
\end{widetext}
Here, $q\geq 1$ is the number of mass bins used, and we have used the following definitions:
\bea
\bar{M}_b \equiv \frac{\sum_j {\cal N}^j_b M_j }{\bar{n}^b_{\rm halo} V},\\
\mu_b \equiv \frac{\bar{n}^b_{\rm halo} \bar{M}_b}{\bar{\rho}}, \\
u^{b}_{\rm AHM}(k) \equiv  \frac{\sum_j {\cal N}^j_b M_j u_{\rm AHM, \bf {k}}^{j}  }{\bar{n}^b_{\rm halo} \bar{M}_b V}.
\eea

Note that, at this level, AHM does not make a prediction for the halo-halo auto-power spectrum, which can be impacted by nonlinear structure formation, and here we rely on N-body simulations to model it.  Furthermore, Equation (\ref{p_hm}) or (\ref{p_matrix}) can be considered as precise definitions of (mean) halo profiles $u^b_{\rm AHM}$ and linear bias $b(\bar{M}_b)$ (by setting $u^b_{\rm AHM}(k=0) \to 0$), and thus make no assumptions about the distribution of matter. 

One may then gain an intuition about the nature of the AHM vs SHM  on large scales, using the linear bias approximation for halo distribution, and the mass function $n(M)$: 
\bea\delta_{m,{\bf k}} = \delta_{L,{\bf k}}+ \sum_j \frac{M^j}{\bar{\rho}}u^j_{\rm AHM,{\bf k}} \exp(i{\bf k\cdot x}_j) \nonumber\\ \simeq \left[1+\frac{1}{\bar{\rho}} \int dM M n(M) b(M) u_{\rm AHM}(k,M)\right] \delta_L(k),\nonumber\\
\label{delta_mass}
\eea
where the fact that $u^b_{\rm AHM}(k=0) \to 0$ guarantees the agreement with linear density predictions on large scales. In contrast, for SHM, we have:
\bea \delta_{m,{\bf k}} \simeq  \left[\frac{1}{\bar{\rho}} \int dM M n(M) b(M) u_{\rm SHM}(k,M)\right] \delta_L(k),\nonumber\\
\label{delta_SHM_mass}
\eea
where $u^b_{\rm SHM}(k=0) \to 1$, and thus an additional condition of $\int dM M n(M) b(M) = \bar{\rho}$ is necessary to recover large scale linear behaviour \footnote{While this condition follows from the assumption that all mass is in haloes, given that $b(M)$ and $n(M)$ are measured from simulations only in a finite mass range, satisfying $\int dM M n(M) b(M) = \bar{\rho}$ requires an additional constraint on extrapolating functions}.

Even then, as we discussed in \cite{Chen:2019wik}, the 1-halo term in SHM spoils the linear behavior of the matter power spectrum on large scales, while the revised equations (\ref{delta_x}) and (\ref{delta_k}) are guaranteed to recover it.

\section{Results and Discussion} \label{sec:results}

\begin{figure}
    \centering
    \includegraphics[width=1.15\linewidth]{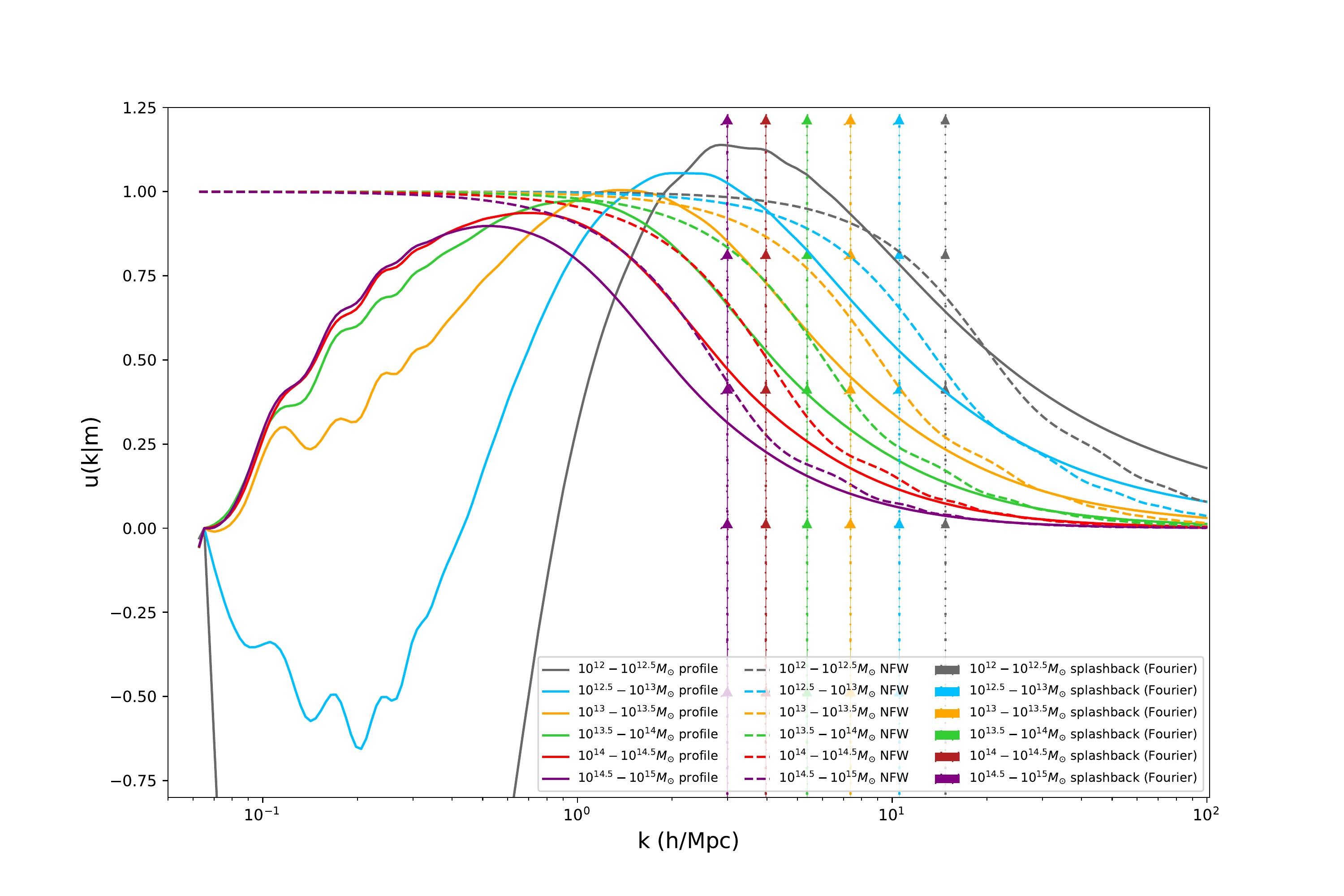}
    \caption[Halo Profile Function in k space]{\footnotesize The halo profile $u(k)$ for all mass bins in Fourier space.  It can be seen here that the peak heights don't differ significantly, but the peak position $k$ increases as the halo mass decreases.  We also see the halo profiles approach the NFW profile at large $k's$/small $r's$, or inside the halo region, as we should expect.  (At smaller k's we see some oscillation in the profile, likely due to approaching the boundaries of the simulation box.)  The splashback radius (arrow lines) is roughly 2$\times$ the k value of the profile peak.}
    \label{fig:r2_profile_plot}
\end{figure}

To find the compensated halo profile for AHM, $u_{\rm AHM}(k|M)$ from Eq.(\ref{p_matrix}), we use simulation data from DarkEmu \cite{Nishimichi:2018etk} for the auto halo power spectrum $P_{hh}$, and halo-mass cross-power spectrum $P_{hm}$.  We then invert Eq.(\ref{p_matrix}) to find the halo profile functions, which are plotted in Figure (\ref{fig:r2_profile_plot}), and compared to the best-fit NFW profiles in the corresponding mass bins. Note that all  $u_{\rm AHM}(k|M)$'s go to zero for $k \to 0$ for compensated haloes, while $u_{\rm NFW}(k) \to 1$ ($u_{\rm NFW}(k)=u_{\rm SHM}(k)$) by construction as NFW haloes are truncated at $r_{\rm virial}$. However, even for larger k's, we find that the simulated profiles don't exactly match the NFW in Fourier space, with the discrepancy being bigger for smaller mass halos.  This is likely due to leakage of structure outside the virial radius into larger k's in the Fourier transform.

\begin{figure*}
    \includegraphics[width=1.0\linewidth]{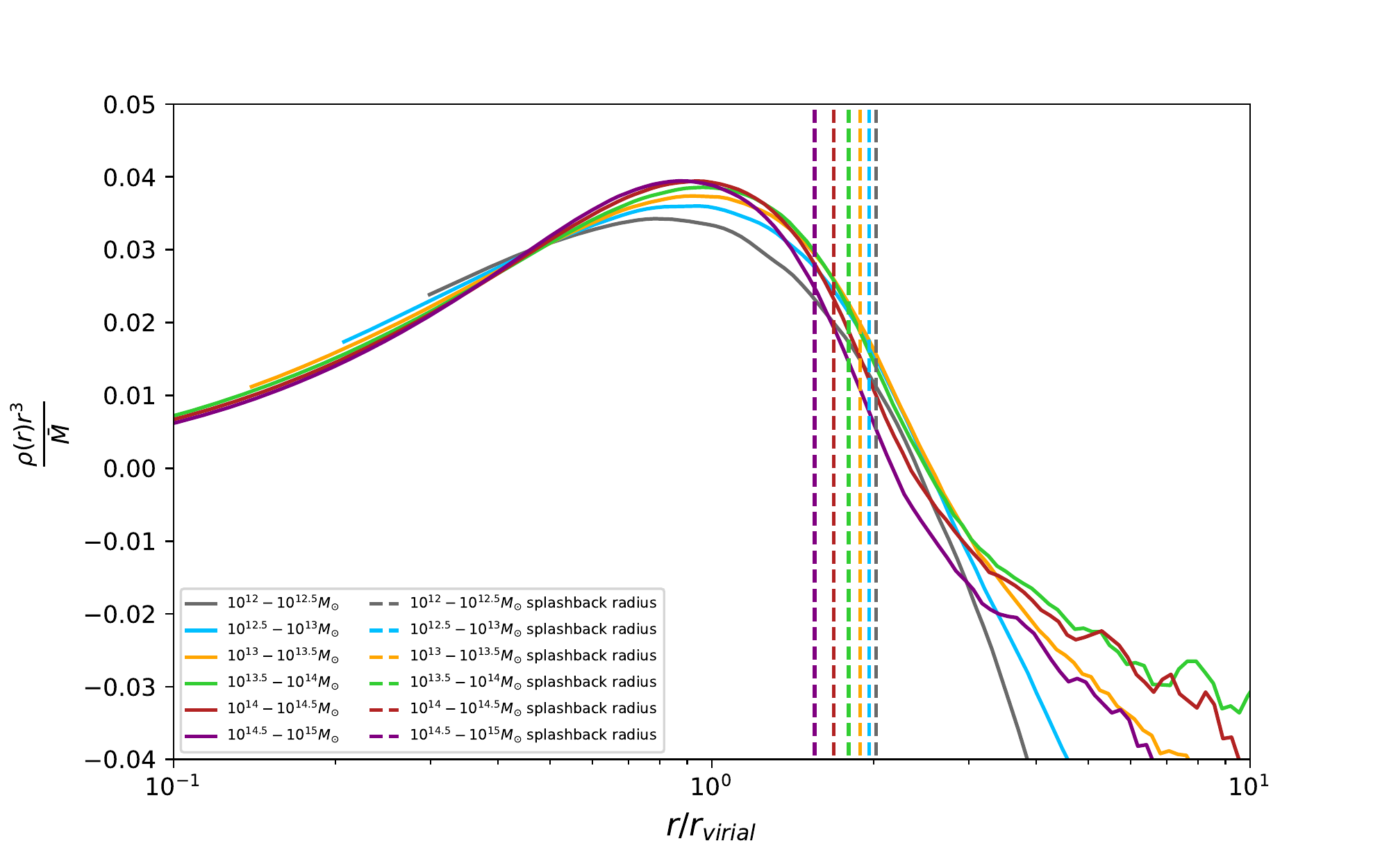}
    \includegraphics[width=1.0\linewidth]{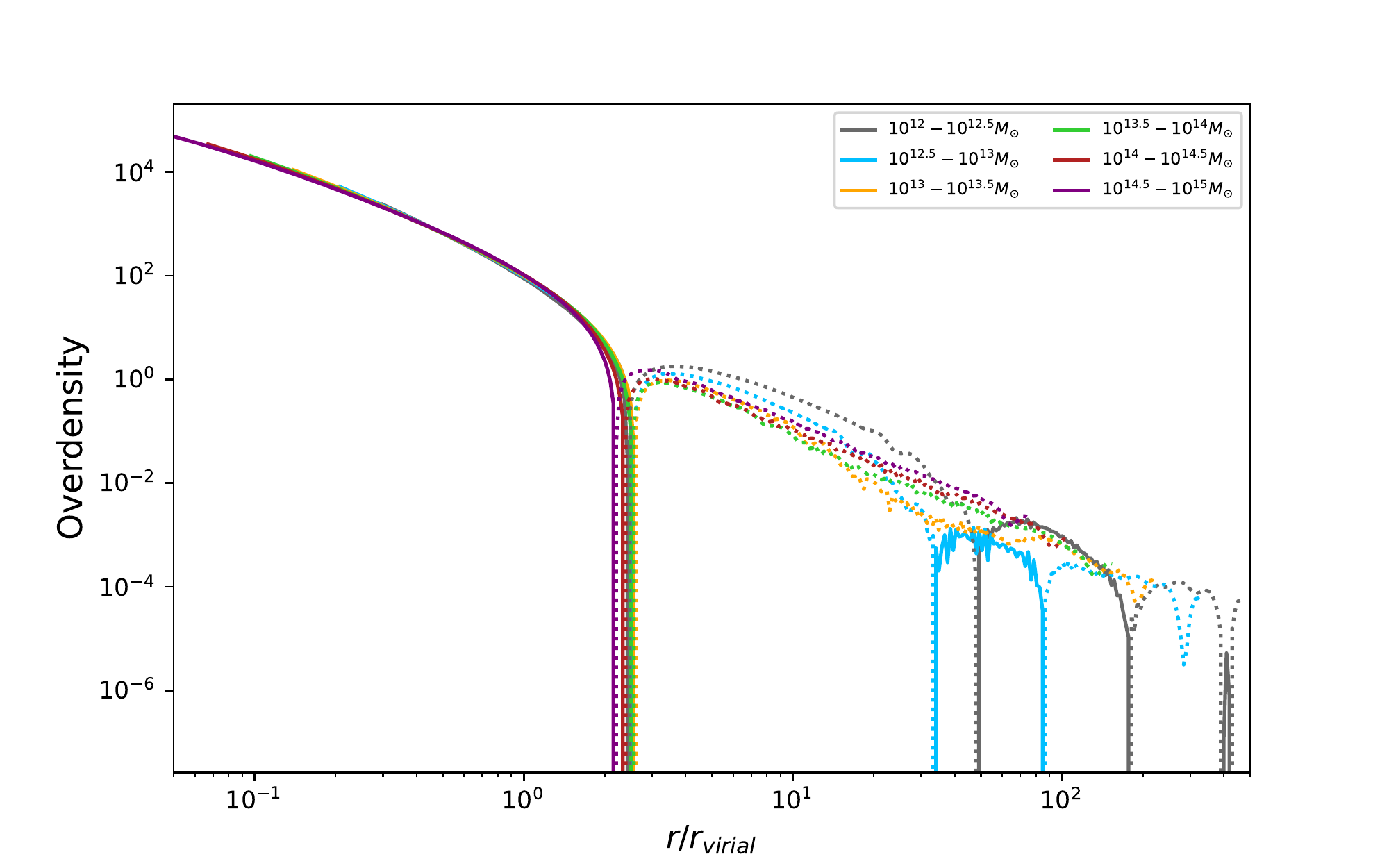}
    \caption[Halo Profile Function in k space]{These figures shows the halo profile $u(r)$ in real space. In panel (a)(top) the profile is multiplied by $r^3$, and plotted as a function of $r/r_{\rm virial}=r/(cr_{scale})$ up to 10 times the virial radius.  This shows that the dimensionless profile peaks around $r_{virial}$ and has near-universal values for all mass bins. The zero crossing happens at distances $20\%-30\%$ outside the splashback radii of the haloes, that are shown for comparison \cite{More:2015ufa}. 
    In panel (b) (bottom), the halo overdensities $\rho(r)/\bar{\rho}$ for each mass bin in real space are shown for our entire radial range, although they are not reliable at very large distances.  The dotted parts show negative overdensity.}  \label{fig:r_profile_overlap}
\end{figure*}

The inferred $u_{\rm AHM}(k|M)$'s can be Fourier transformed to find real-space halo profiles, $u(r) = \rho(r)/M$. Figure (\ref{fig:r_profile_overlap}) shows these real-space profiles $u(r)\times r^3$ as a function $r/r_{\rm virial}$. As expected, what we find is that the compensated profile goes negative in the outskirts, just outside the splash-back radii \cite{More:2015ufa} of the haloes.

In the inner regions, the dimensionless profile $r^3\times u(r)$ has a near-universal positive maximum around $r = r_{\rm virial}$ (with values ranging from 0.034 to 0.039, for smallest to largest masses).    The profile is also mostly independent of the minimum mass bin of the halo with a key exception - for results to be reasonable, the smallest mass bin should be $\gtrsim 10^{12} M_\odot$ - which is set by the minimum mass of haloes resolved in DarkEmu simulations.  

\begin{figure}
    \centering    \includegraphics[width=1.1\linewidth]{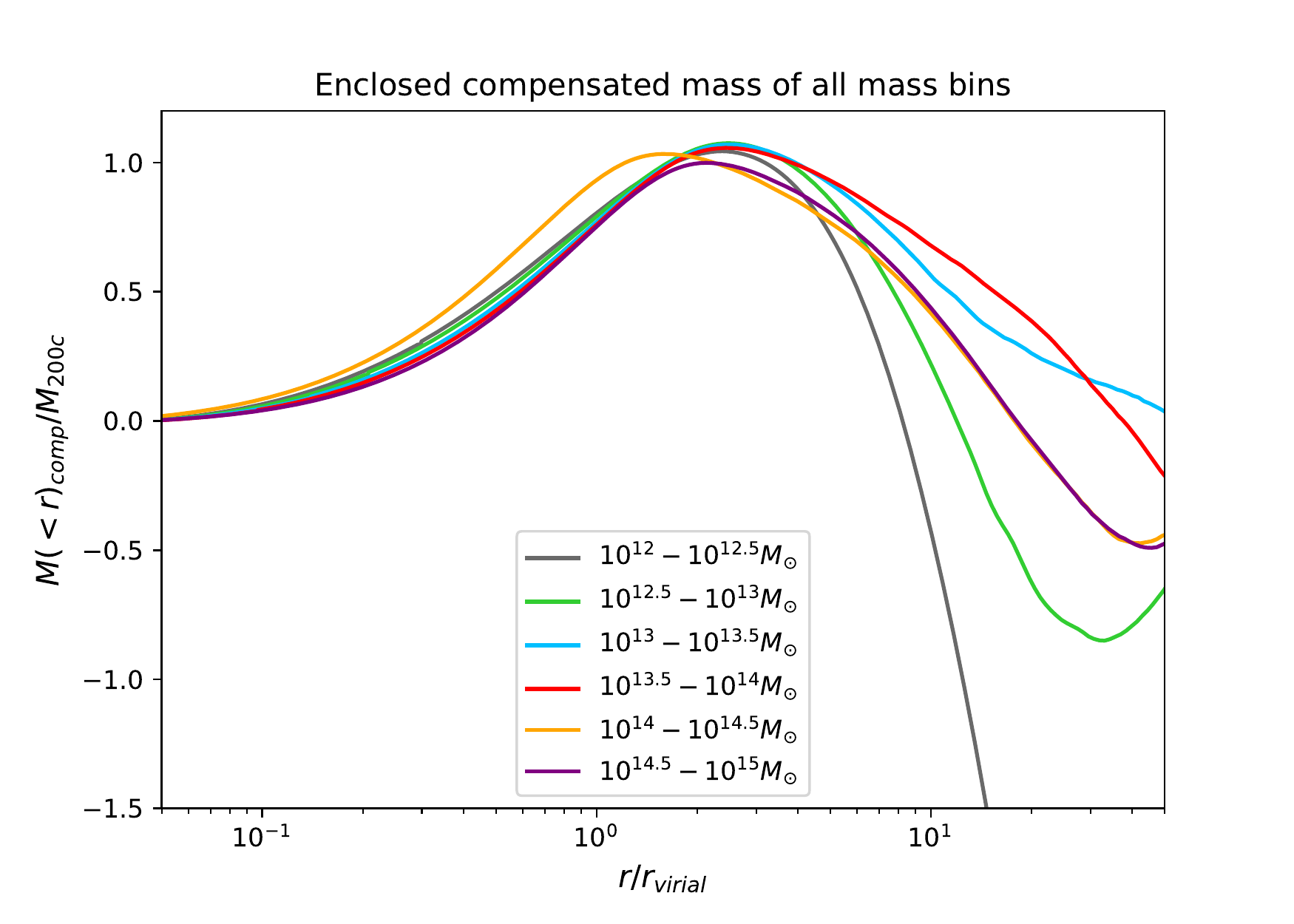}
    \caption[Enclosed mass in k space]{\footnotesize The integrated enclosed mass found in real space using the compensated halo profile.  Similar to the halo profiles, the peak height of the enclosed masses for different mass bins are also almost the same - the peak is independent of halo mass.}
    \label{fig:mass_enclosed}
\end{figure}

Figure (\ref{fig:mass_enclosed}) shows the enclosed mass (in units of virial mass), that results from integrating Figure (\ref{fig:r_profile_overlap})\footnote{We use the NFW profile to integrate mass within radii below the spatial resolution of the DarkEmu}. The radii of the peaks of these mass profiles would thus coincide with zero crossings of the density. We further notice that the maximum compensated masses (masses of compensated halos) are nearly the same as uncompensated virial mass, although the latter is defined at a smaller radius. 

\begin{figure}
    \includegraphics[width=1.\linewidth]{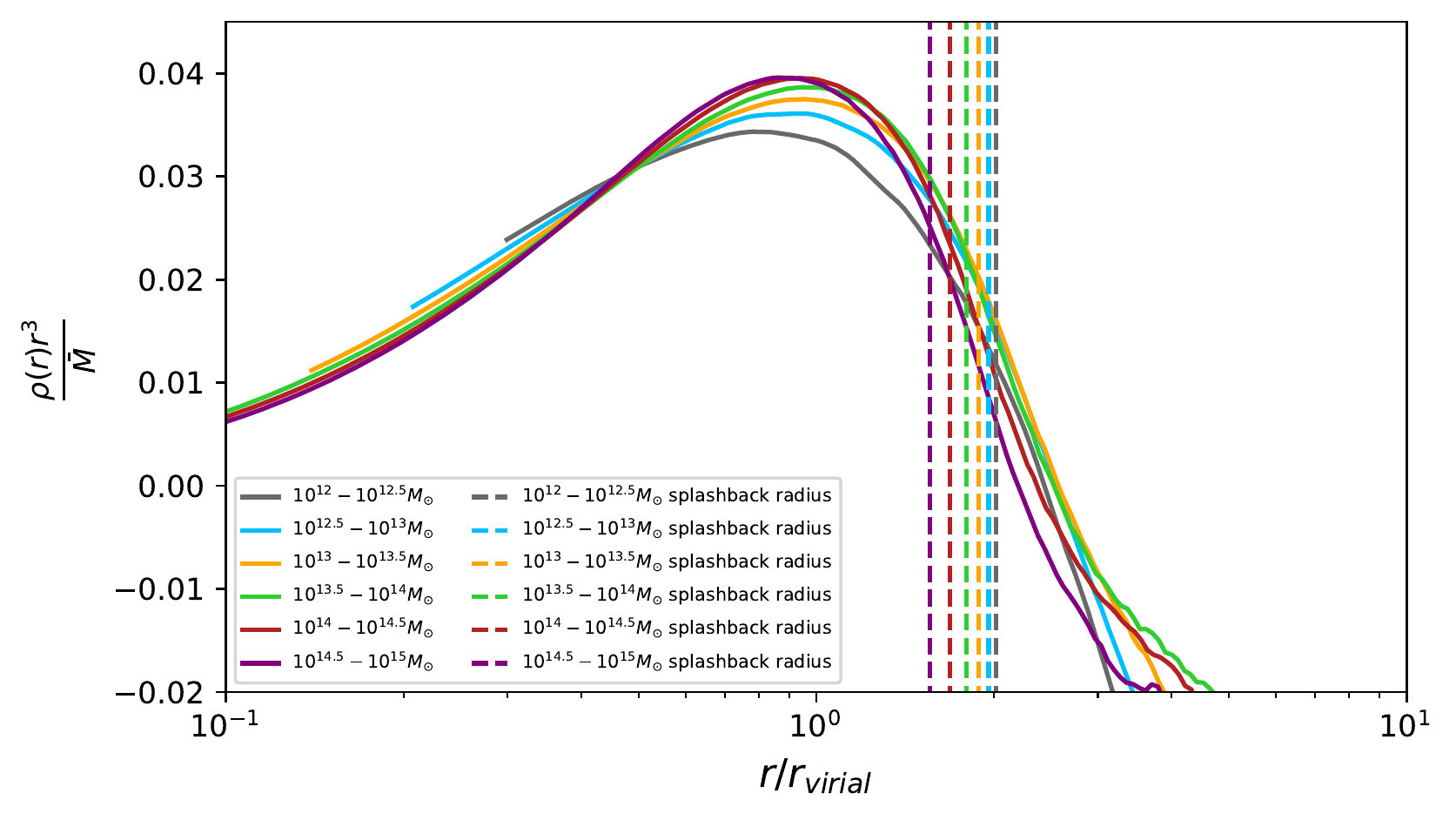}     \includegraphics[width=1.\linewidth]{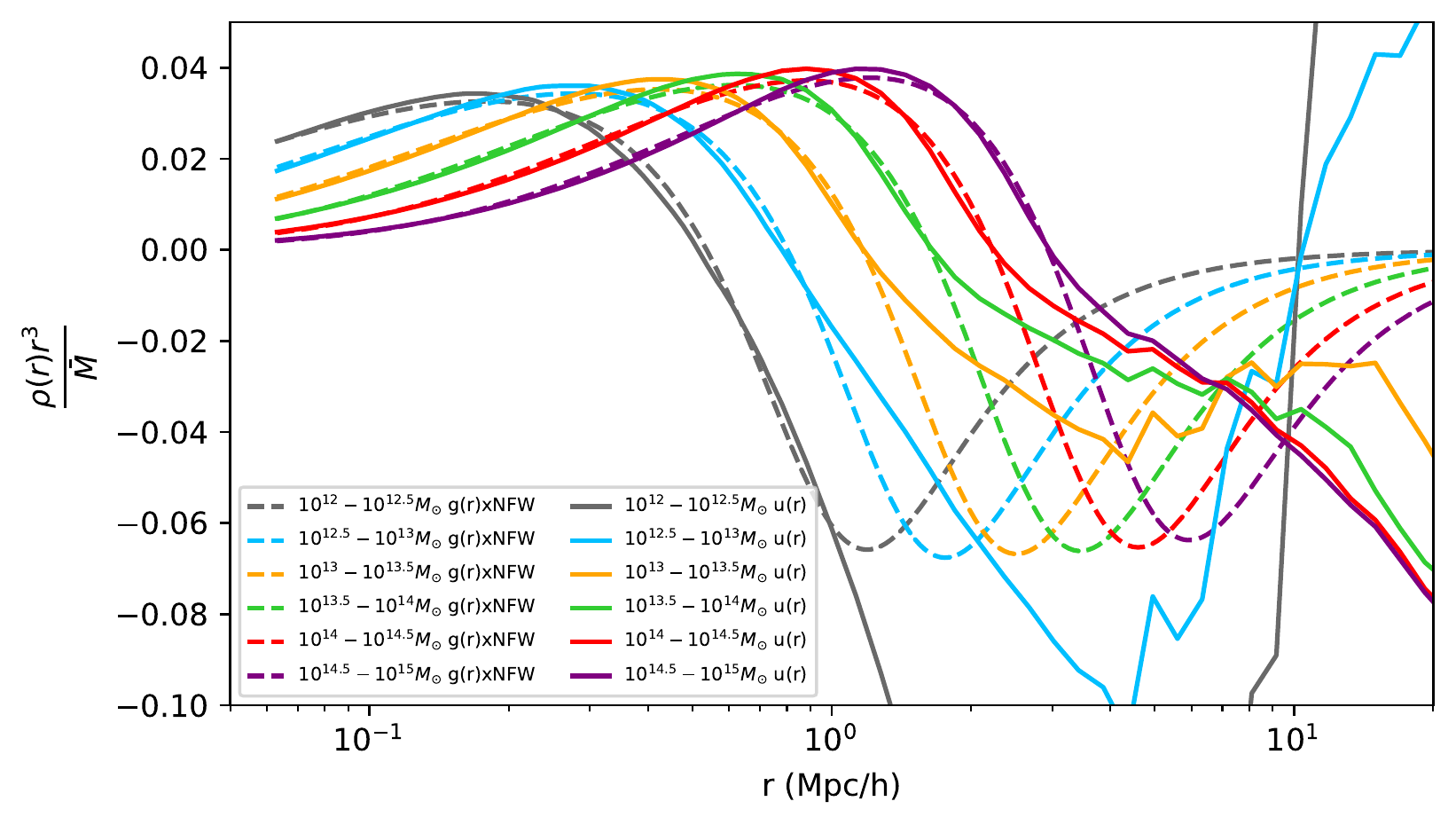}
    \caption[Halo Profile Function in k space]{(a) (top) \footnotesize The $u(r)$ function times $r^3$ in real space compared to the NFW profile.  It can be seen that at small r's, the two start to approach each other, as we would expect, since NFW describes halo density at small radii inside the halo (this likely deviates near the halo centre due to baryonic effects, but we don't go to that scale here).  (b) (bottom) The $u(r)$ ($u(k|m)$ function Fourier transformed into real space) fitted to a function with two free parameters (see Table 1).  The fit works up to around 1-4Mpc/h, after which the Fourier transform noise takes over.}
    \label{fig:u_NFW}
\end{figure}

\begin{table}
 \begin{tabular}{||c | c | c | c | c | c | c ||} 
 \hline
 Mass bin & $\alpha$ & $\beta$ & $r_{zc}$ & $\frac{ r_{zc} }{ r_{vir} }$ & $r_{peak}$ & $\frac{ r_{peak} }{ r_{vir} }$ \\ [1.5ex] 
 \hline
 $10^{12-12.5}M_{\odot}$ & 0.80 & $4.0\times 10^{-7}$ & 0.53 & 2.5 & 0.18 & 0.84 \\ 
 \hline
 $10^{12.5-13}M_{\odot}$ & 0.79 & $6.9\times 10^{-7}$ & 0.80 & 2.6 & 0.28 & 0.91 \\ 
 \hline
 $10^{13-13.5}M_{\odot}$ & 0.77 & $1.6\times 10^{-6}$ & 1.2 & 2.6 & 0.42 & 0.93 \\
 \hline
 $10^{13.5-14}M_{\odot}$ & 0.76 & $3.5\times 10^{-6}$ & 1.6 & 2.5 & 0.62 & 0.96 \\ 
 \hline
 $10^{14-14.5}M_{\odot}$ & 0.76 & $8.3\times 10^{-6}$ & 2.2 & 2.3 & 0.88 & 0.94 \\ 
 \hline
 $10^{14.5-15}M_{\odot}$ & 0.76 & $2.0\times 10^{-5}$ & 2.9 & 2.1 & 1.2 & 0.91 \\  [1.5ex] 
 \hline
\end{tabular}
\caption{\footnotesize Best fit parameters at z=0 (present day) \footnote{All the results presented here are at z=0.  There could be redshift dependence on halo bias and distributions that would affect these parameters, but this is currently not well understood beyond the scope of this work.  We would expect $r_{peak}$ to decrease as redshift increases though, as haloes would have less mass at earlier formation stages. } for Eq. (\ref{eqn:gr_fit}) shown by $\alpha$ and $\beta$.  We also report the best-fit zero-crossing radius $r_{zc}$, the ratio of the zero-crossing radius and the virial radius $r_{zc}/r_{vir}$, the peak radius $r_{peak}$ (where $u(r)r^3$ is at a maximum), as well as the ratio of the peak radius to the virial radius.}
\label{tab:params}
\end{table}

In Figure \ref{fig:u_NFW}, we fit our measured halo profiles in real space to a function of the form
\beq
M\times u_{\rm AHM}(r|M) = g(r)\times \rho_{\rm NFW}(r),
\label{eqn:new_fit}
\eeq 
where $\rho_{\rm NFW}$ is the best-fit NFW profile for the corresponding mass bin,  and $g(r)$ is a simple fitting function (with two free parameter $\alpha$ and $\beta$) that approaches $\alpha$ at small $r$, but goes negative at $r> r_{zc}$, and is constructed such that $g \times \rho_{\rm NFW}$ is compensated (see \ref{sec:appendix} for the full form of $g(r)$).

Table \ref{tab:params} shows our numerical best fits for $\alpha$ and $\beta$ for each mass bin. We also see that, as we noted above, the peak of $r^3\times u(r|M)$ is always within a few percent of the virial radius, while the zero crossing happens at $r_{zc}$, between 2-3 times the virial radius (or 1.2-1.3 times the splashback radius, see Figure \ref{fig:r_profile_overlap}).

\begin{figure}
    \includegraphics[width=1.\linewidth]{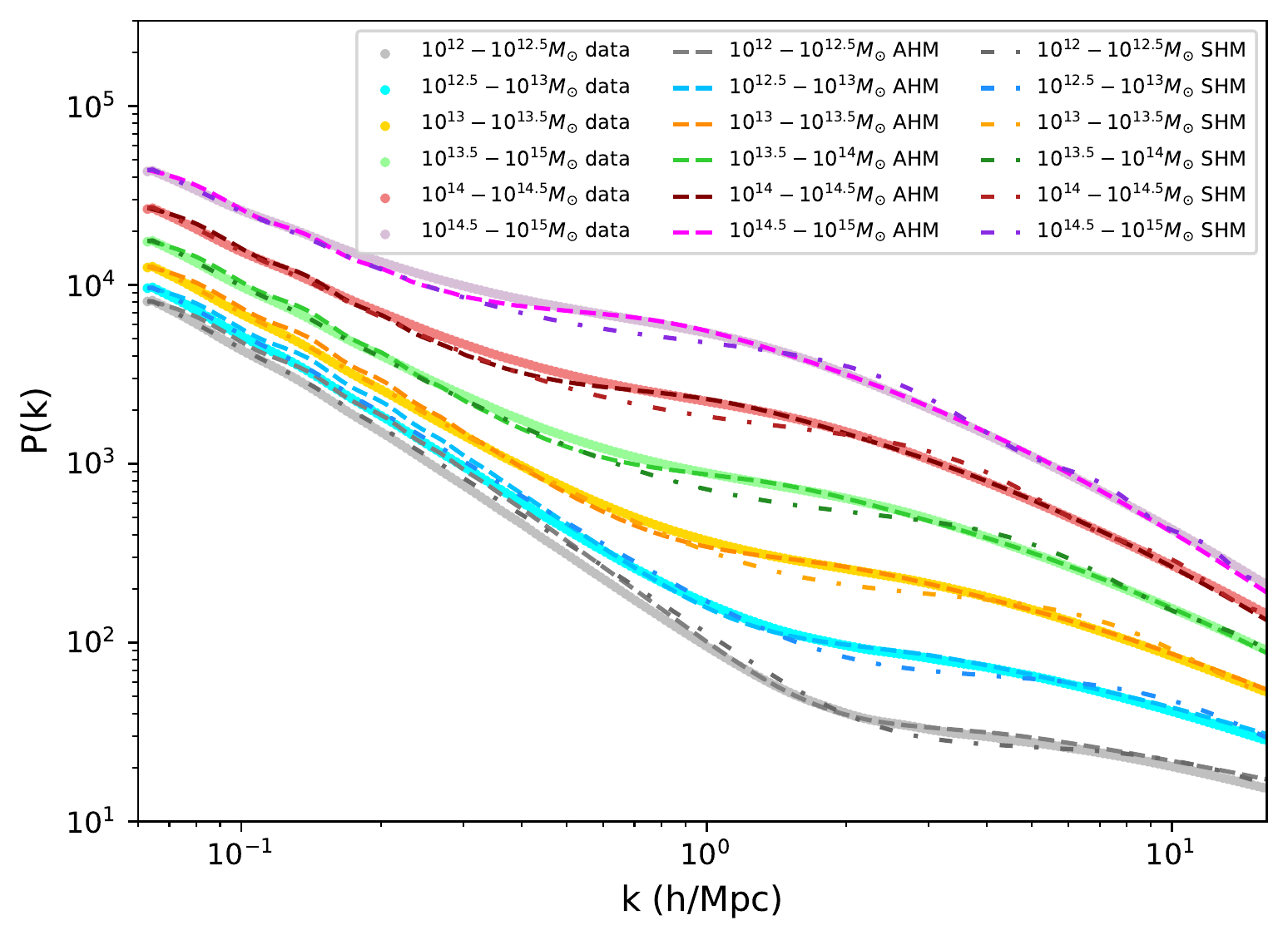}
    \includegraphics[width=1.\linewidth]{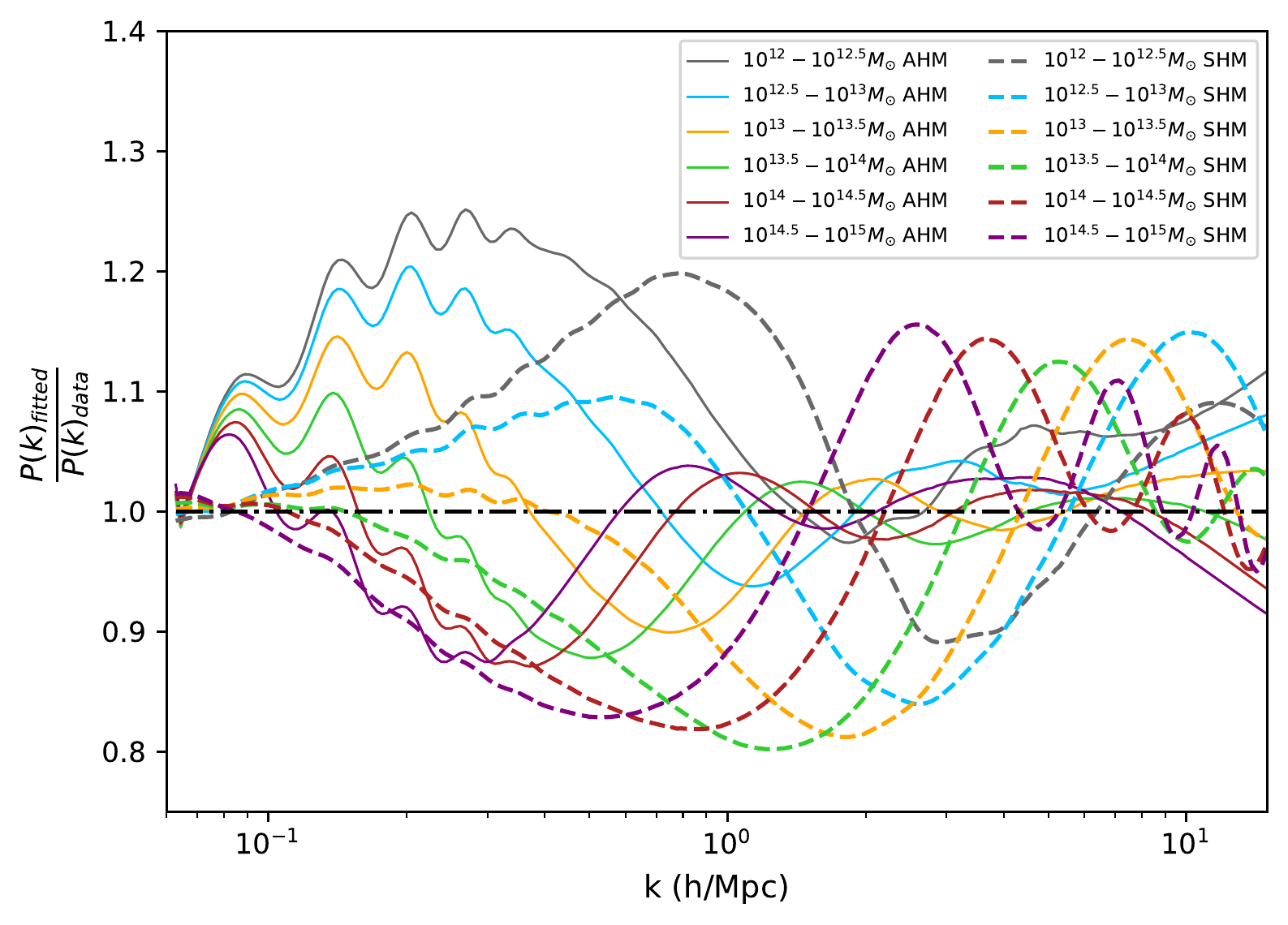}
    \includegraphics[width=1.1\linewidth]{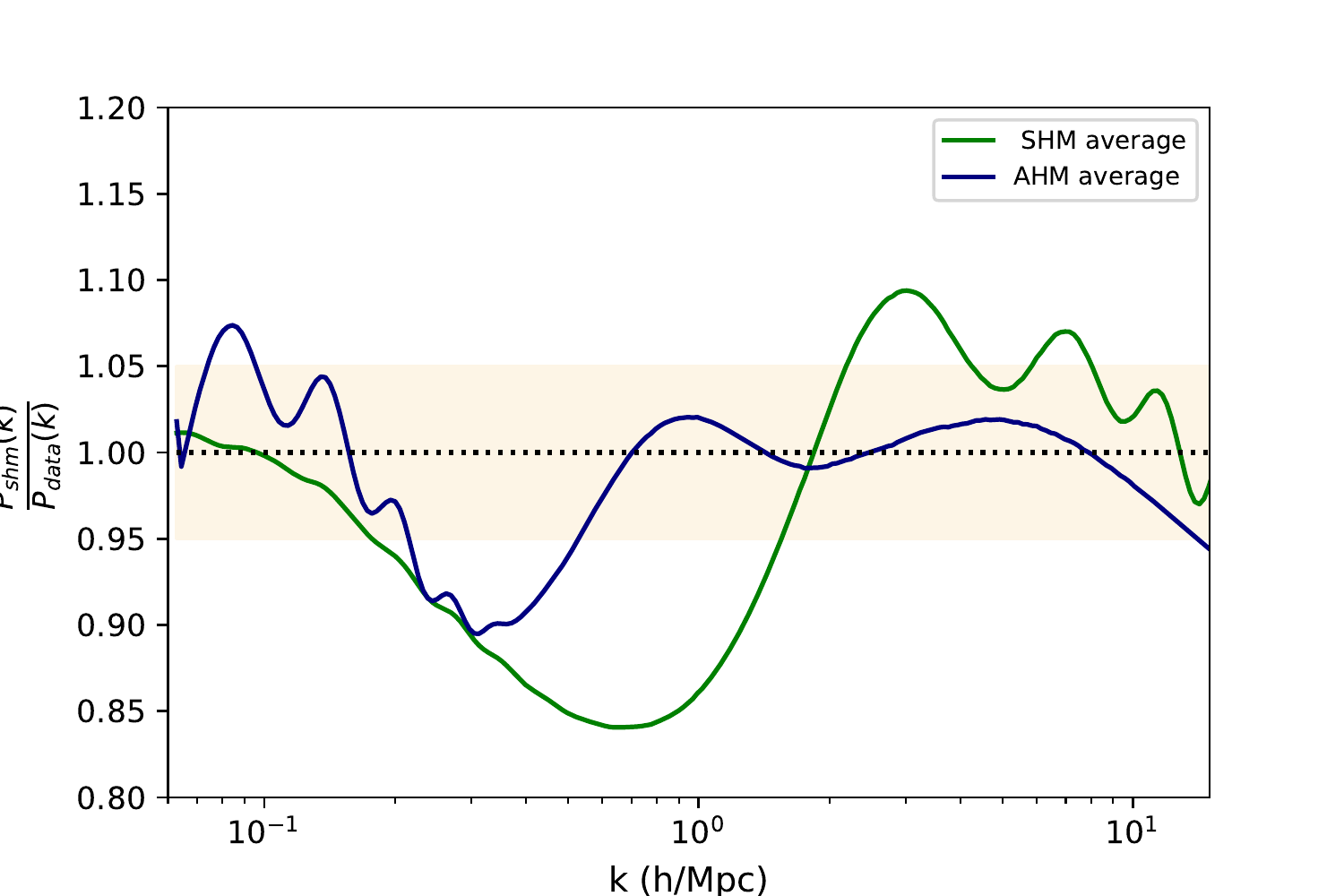}
    
    \caption[Fits to Halo-matter power spectrum]{\footnotesize (a) (top) Comparison of matter-halo cross-power spectra, between DarkEmu (dots), our fitted compensated profiles (Equation \ref{eqn:new_fit}), and standard halo model profiles  We can see that at intermediate k's they all mostly follow each other, which is what we would expect.
    (b) (middle) The ratios of our compensated fit and standard halo model to DarkEmu cross-power spectra for each mass bin.  The fit works better for larger mass bins, where the DarkEmu simulation data is more robust. Moreover, note that the fit in Equation (\ref{eqn:new_fit}) is only done in real space and radii $r< 3~ h^{-1}{\rm Mpc}$, which is why the behavior at smaller $k$'s is not well-modelled by the fit. Furthermore, k $\ll$ 1.0 hMpc$^{-1}$ are likely not well-resolved in simulations.
    (c) (bottom) The ratios of the SHM and AHM models to the DarkEmu cross-power spectra, averaged over the mass bins from $10^{13}-10^{14.5} \odot$.  At small k's SHM does slightly better than our amended fit, but overall across the k range our amended fit does better.  The light orange band shows the error range within 5$\%$ of the simulation data.}
    \label{fig:pk_fit_all}
\end{figure}

Using this fitted halo profile to account for compensation, we re-calculated the power spectrum to see how well it matched the data from DarkEmu.  The results are shown in the top and bottom plots of Figure \ref{fig:u_NFW}, which shows that our theoretical fit is capable of matching the simulation data within the range where the halo profile is not noise dominated (both inside and outside the halo regions).

From the Figures \ref{fig:u_NFW} (a) and (b), it can be seen that it is possible to fit the cross halo-matter power spectrum using a compensated halo profile term as well.  While, so far our fitted $\beta$ parameters do not have a straightforward physical interpretation ($\alpha$ is only a simple normalization), our exercise shows that both the matter auto spectrum, as well as the matter-halo cross power spectra, can be modelled with compensated halo profiles. 

Figure \ref{fig:pk_fit_all} compares the resulting cross-power spectra from Equation (\ref{eqn:new_fit}) with DarkEmu, showing reasonable agreement at small scales (used in the fit), but modest $20-30\%$ disagreement for larger scales.     
For practical applications, the accuracy of the fit can be improved if more parameters are added, but that is beyond the scope of this study.  The main point we want to show is that compensation is an important part of the halo model in both the matter auto power spectrum and the halo-matter cross spectrum, as we can see that adding compensation provides a better fit to simulation data overall across a wider range of k's.  The oscillations around the halo boundary cut-off in the standard halo model are also avoided, as we extropolate the profile beyond the virial radius.

Having a physically motivated model for the power spectrum - both matter auto and halo-matter cross spectra - can help us predict power at k values beyond what simulations can currently resolve, at either very high or very low k's.  This can be applied to lensing observables as well (see Figure \ref{fig::lensing_power} in \ref{sec:lensing}), since a compensated halo model will predict observed lensing power more accurately at small k's/large radii, where the standard halo model overpredicts power.  Also, it can be seen from Figure \ref{fig:pk_fit_all} that a compensated model still is a better fit at around k$\sim$1h/Mpc, near the halo boundaries, where SHM usually does not model well.  Thus, having a semi-analytical formalism for cross halo-matter power spectra can be useful when making predictions on a wide scale beyond what simulations can currently produce \cite{Asgari:2023mej} and SHM can accurately predict.

\section{Conclusion and Future Prospects} \label{sec:conclusion}

In this paper, we built upon our previous study \cite{Chen:2019wik}, to show that compensation -- introduced as an important amendment of the halo model to respect physical conservation laws -- can be applied to model both the auto-matter power spectrum and halo-matter cross-power spectra.  Compensated halo profiles still match NFW in the inner halo regions, but now also take into account underdense regions (or voids) in the outskirts of cosmological haloes.  The new compensated (dimensionless) profiles show a near-universal behaviour out to the splashback radius, independent of the halo mass. What is notable is that the dimensionless profiles peak at the virial radius and the profiles of different halo masses have near-universal peak values, while the maximum compensated mass coincides with the (uncompensated) virial mass of the haloes.  This new halo profile can also be fitted numerically by introducing two free parameters, which we report in Table \ref{tab:params}.  The physical interpretation the fitting parameters are currently beyond the scope of this paper, and we leave them for future work. More precise fits (especially outside $r \gtrsim 10.0 h^{-1} {\rm Mpc}$) can be obtained by using more parameters in the fitting function, though the goal here is not to obtain a model of extremely high precision.  Rather, we want to show halo profiles have similar peak heights across different mass bins and these profiles can be fitted using amendments (the result of adding halo compensations). This will be an important step to compare the predictions of the amended halo profile for weak lensing power and kinetic Sunyaev-Zel'dovich (kSZ) observations, where mass and momentum conservation may play an important role in constraining physical possibilities on intermediate and large scales.  Semi-analytic formalisms are currently important for this, as simulations have finite resolution and k-modes.

Other factors not taken into account here are sub-structure of halos, and halo assembly bias \cite{Dalal:2008zd} - for the latter, we did try different bias models (\cite{Tinker:2008ff}, \cite{Mead:2020qdk}), but it did not yield any significant changes in the results.  Effects that assembly bias, halo sub-structure, or filaments may further have on the compensated halo profiles are interesting future prospects to explore. In particular, the extent to which AHM can be used (or adapted) to model the matter/halo bi-spectra (or 3-point correlation function) is another interesting direction, as it may probe the covariance of compensated profiles at large distances; an important step for kSZ studies will be the determination of compensated momentum profiles (due to gas infall in halo outskirts) in the hydrodynamical simulations.  In addition, the mass clustering in haloes will affect the rotation curves of the galaxies residing inside these haloes \cite{Dai:2022had}, which can provide evidence for different cosmological models, such as $\Lambda$CDM vs. MOND \cite{Dai:2022had}.  All these effects are interesting future paths of exploration for dark matter halo profiles.


\begin{acknowledgments}
We would like to thank Jenny Wagner, Vincent Desjacques, Alexander Mead, Fabian Schmidt, and Dejan Stojkovic, as well as the anonymous reviewer, for helpful discussions, suggestions, and comments.

We would also like to thank Ryuichi Takahashi, Masanori Sato, Takahiro Nishimichi, Atsushi Taruya, and Masamune Oguri for letting us use their large scale simulation data for the matter auto-power spectra, as well as the DarkEmu team for their halo-matter cross-power spectra simulation data \cite{Nishimichi:2018etk}. For calculating linear and numerical nonlinear power spectra for comparison with our model, we used the CAMB package in Python \cite{Lewis:1999bs, Peacock:2000qk, Takahashi:2012em}.

AC is funded by the University of Waterloo and Perimeter Institute for Theoretical Physics. NA is funded by the University of Waterloo, the National Science and Engineering Research Council of Canada (NSERC) and the Perimeter Institute for Theoretical Physics.
Research at Perimeter Institute is supported by the Government of Canada through Industry Canada and by the Province of Ontario through the Ministry of Economic Development \& Innovation. 
\end{acknowledgments}

\appendix

\section{Explicit profiles}
\label{sec:appendix}
The NFW function in Fourier space is given by:
\begin{widetext}
\beq
u_{\rm NFW}(k|M) = \frac{4\pi {\rho}_s r_s^3}{M}\left\{ \sin(kr_s) \big({\rm Si}[(1+c)kr_s] - {\rm Si}(kr_s)\big) + \cos(kr_s)\big( {\rm Ci}[(1+c)kr_s] - {\rm Ci}(kr_s)\big) - \frac{\sin(ckr_s)}{(1+c)kr_s} \right\},
\label{shm_density}
\eeq
\end{widetext}
and this is the equation that our fitting function tends to in the limit $r \ll r_{\rm virial}$.

The fitting function $g(r)$ in Eq. \ref{eqn:new_fit} is of the form
\begin{widetext}
\beq
g(r) = \alpha \bigg(1 - \frac{\frac{\sqrt{\beta} (-4(1+\beta) + \beta^{1/4}(2\sqrt{2}-3\beta^{1/4}+4\sqrt{2\beta}-2\sqrt{2}\beta + \beta^{5/4})\pi +(-1+3\beta)ln(\beta))}{\pi(1-2\sqrt{2}\beta^{1/4}+4\sqrt{2}\beta^{3/4}-3\beta+2\sqrt{2}\beta^{5/4}) + \sqrt{\beta}(-4(1+\beta) + ln(\beta)(\beta-3))} (r/r_{scale})^2}{1 + \beta (\frac{r}{r_{scale}})^4 } \bigg),
\label{eqn:gr_fit}
\eeq
\end{widetext}

where $\alpha$ and $\beta$ are parameters fitted by least square fitting, listed in Table \ref{tab:params}.  This function is designed to be of the form
\beq 
g(r) \propto \frac{1 - s(\beta) r^2}{1 + \beta r^4},
\eeq 
where $s(\beta)$ is found by integrating $g(r) \times \rho_{\rm NFW}$ from 0 to infinity and setting the integral to 0.  As a result, the final halo profile is compensated.

\begin{figure}[H]
    \centering
    \includegraphics[width=\linewidth]{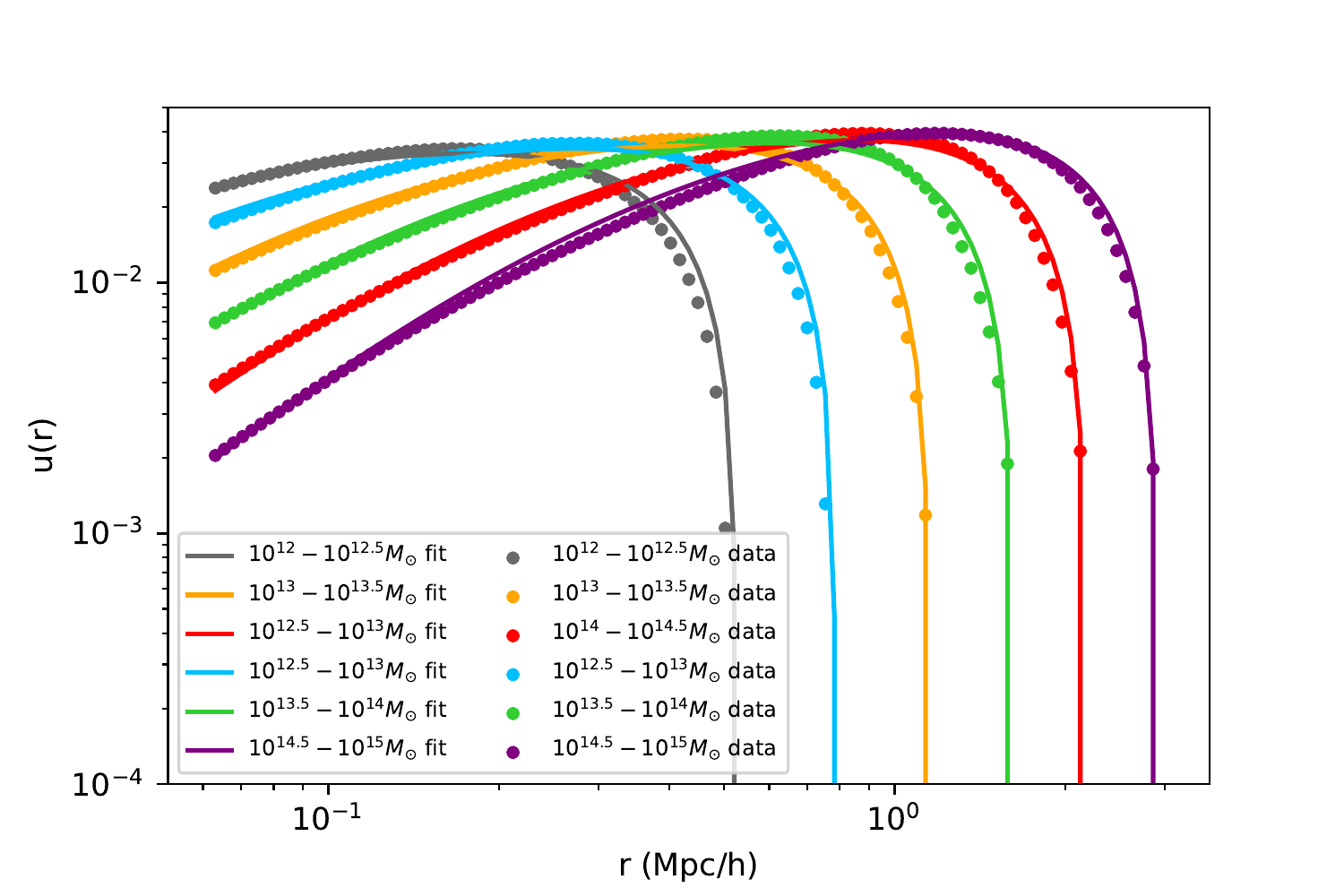}
    \caption[Fits to Halo-matter power spectrum]{\footnotesize A comparison of the simulation halo profile (scatter points) and our compensated profile $u(r)$ in Fourier space, zoomed in at smaller r's.}
\label{fig:ur_zoom}
\end{figure}

\begin{figure}[H]
    \centering
    \includegraphics[width=\linewidth]{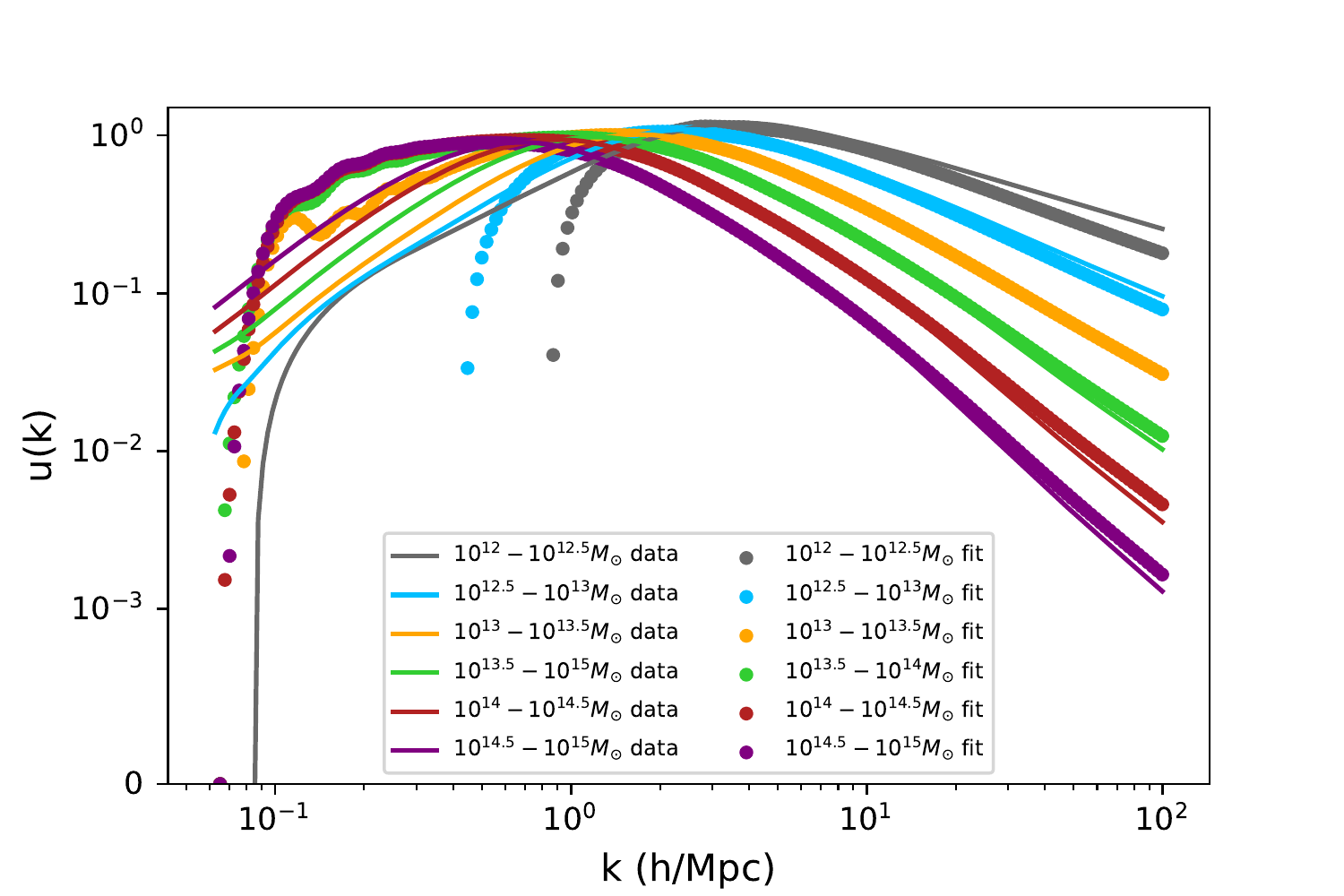}
    \caption[Fits to Halo-matter power spectrum]{\footnotesize A comparison of the simulation halo profile (scatter points) and our compensated profile $u(k|m)$ in Fourier space, zoomed in at smaller k's.  It can be seen here that even though the compensated profile matches the data fairly well in real space at large r's, the same isn't true for small k's, so the relation between the two isn't exactly proportional due to k-mode mixing.}
\label{fig:uk_limit}
\end{figure}

\section{Lensing Power}
\label{sec:lensing}
\begin{figure} [H]
    \centering    \includegraphics[width=1.05\linewidth]{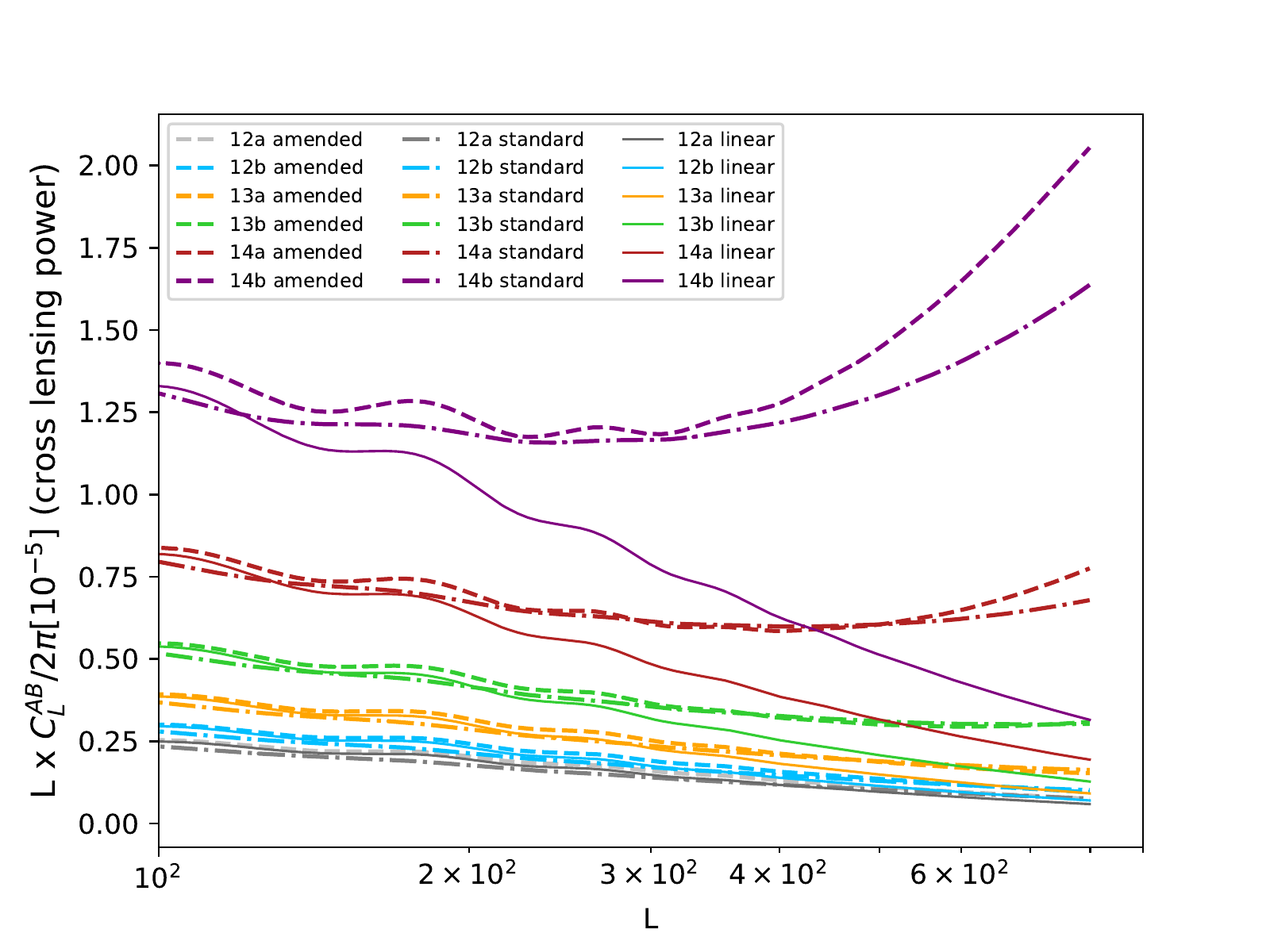}
    \includegraphics[width=1.\linewidth]{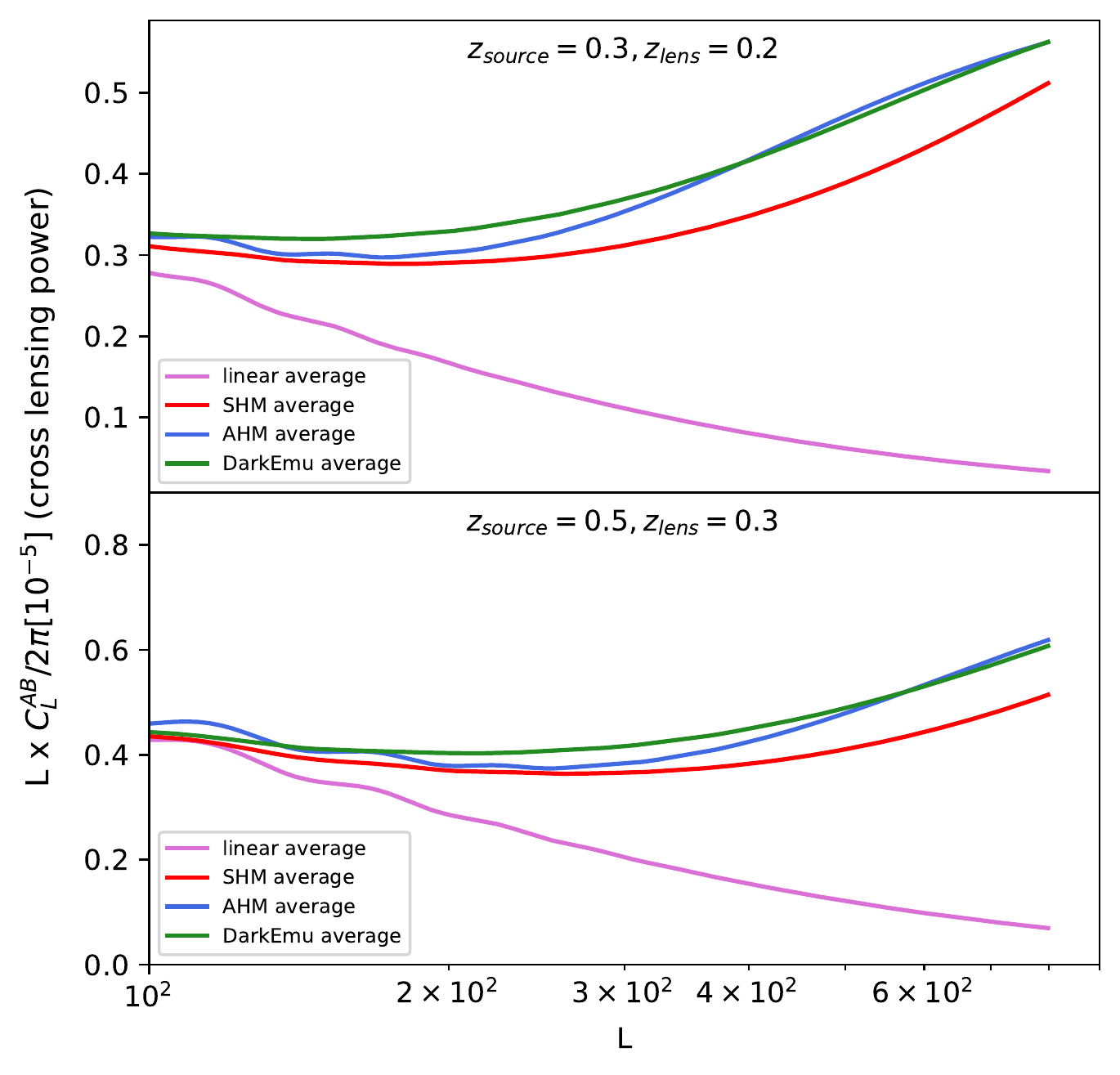}
    \caption{\footnotesize a) (top) Lensing power of the standard halo model, our compensated model, and linear power as a function of angular separation L.  The difference is more significant for larger mass bins, where our compensation has the most effect.
    b) (bottom) Average lensing power of the simulated power from DarkEmu, the standard halo model, our compensated model, and linear power as a function of angular separation L at different redshifts.  We can see from the graph that the compensated/amended model does a better job of fitting the simulated lensing power (DarkEmu) overall throughout the L range. }
    \label{fig::lensing_power}
\end{figure}

A potential observable for the power spectra of large scale structure is the cross halo-matter lensing power (\cite{DES:2021wwk}, \cite{Heymans:2020gsg}), which can be calculated using the Limber approximation \cite{LoVerde:2008re}.  The lensing power for the halo-matter cross power spectra (standard halo model, compensated model, and linear power) can be seen in Figure \ref{fig::lensing_power}.  We see that the nonlinear lensing power follows the data in Figures 2 $\&$ 3 in \cite{Heymans:2020gsg} more closely, which is what we would expect to see.  The amended (compensated) model also follows the plateau pattern in \cite{Heymans:2020gsg} slightly better than the standard model at large L's, so it could potentially be a more accurate model in a bigger observational region of the sky.

\bibliography{references}

\begin{thebibliography}{24}%
\makeatletter
\providecommand \@ifxundefined [1]{%
 \@ifx{#1\undefined}
}%
\providecommand \@ifnum [1]{%
 \ifnum #1\expandafter \@firstoftwo
 \else \expandafter \@secondoftwo
 \fi
}%
\providecommand \@ifx [1]{%
 \ifx #1\expandafter \@firstoftwo
 \else \expandafter \@secondoftwo
 \fi
}%
\providecommand \natexlab [1]{#1}%
\providecommand \enquote  [1]{``#1''}%
\providecommand \bibnamefont  [1]{#1}%
\providecommand \bibfnamefont [1]{#1}%
\providecommand \citenamefont [1]{#1}%
\providecommand \href@noop [0]{\@secondoftwo}%
\providecommand \href [0]{\begingroup \@sanitize@url \@href}%
\providecommand \@href[1]{\@@startlink{#1}\@@href}%
\providecommand \@@href[1]{\endgroup#1\@@endlink}%
\providecommand \@sanitize@url [0]{\catcode `\\12\catcode `\$12\catcode
  `\&12\catcode `\#12\catcode `\^12\catcode `\_12\catcode `\%12\relax}%
\providecommand \@@startlink[1]{}%
\providecommand \@@endlink[0]{}%
\providecommand \url  [0]{\begingroup\@sanitize@url \@url }%
\providecommand \@url [1]{\endgroup\@href {#1}{\urlprefix }}%
\providecommand \urlprefix  [0]{URL }%
\providecommand \Eprint [0]{\href }%
\providecommand \doibase [0]{http://dx.doi.org/}%
\providecommand \selectlanguage [0]{\@gobble}%
\providecommand \bibinfo  [0]{\@secondoftwo}%
\providecommand \bibfield  [0]{\@secondoftwo}%
\providecommand \translation [1]{[#1]}%
\providecommand \BibitemOpen [0]{}%
\providecommand \bibitemStop [0]{}%
\providecommand \bibitemNoStop [0]{.\EOS\space}%
\providecommand \EOS [0]{\spacefactor3000\relax}%
\providecommand \BibitemShut  [1]{\csname bibitem#1\endcsname}%
\let\auto@bib@innerbib\@empty
\bibitem [{\citenamefont {Chen}\ and\ \citenamefont
  {Afshordi}(2020)}]{Chen:2019wik}%
  \BibitemOpen
  \bibfield  {author} {\bibinfo {author} {\bibfnamefont {A.~Y.}\ \bibnamefont
  {Chen}}\ and\ \bibinfo {author} {\bibfnamefont {N.}~\bibnamefont
  {Afshordi}},\ }\href {\doibase 10.1103/PhysRevD.101.103522} {\bibfield
  {journal} {\bibinfo  {journal} {Phys. Rev. D}\ }\textbf {\bibinfo {volume}
  {101}},\ \bibinfo {pages} {103522} (\bibinfo {year} {2020})},\ \Eprint
  {http://arxiv.org/abs/1912.04872} {arXiv:1912.04872 [astro-ph.CO]}
  \BibitemShut {NoStop}%
\bibitem [{\citenamefont {Navarro}\ \emph {et~al.}(1996)\citenamefont
  {Navarro}, \citenamefont {Frenk},\ and\ \citenamefont
  {White}}]{Navarro:1995iw}%
  \BibitemOpen
  \bibfield  {author} {\bibinfo {author} {\bibfnamefont {J.~F.}\ \bibnamefont
  {Navarro}}, \bibinfo {author} {\bibfnamefont {C.~S.}\ \bibnamefont {Frenk}},
  \ and\ \bibinfo {author} {\bibfnamefont {S.~D.~M.}\ \bibnamefont {White}},\
  }\href {\doibase 10.1086/177173} {\bibfield  {journal} {\bibinfo  {journal}
  {Astrophys. J.}\ }\textbf {\bibinfo {volume} {462}},\ \bibinfo {pages} {563}
  (\bibinfo {year} {1996})},\ \Eprint {http://arxiv.org/abs/astro-ph/9508025}
  {arXiv:astro-ph/9508025 [astro-ph]} \BibitemShut {NoStop}%
\bibitem [{\citenamefont {Ludlow}\ and\ \citenamefont
  {Angulo}(2017)}]{Ludlow:2016qow}%
  \BibitemOpen
  \bibfield  {author} {\bibinfo {author} {\bibfnamefont {A.~D.}\ \bibnamefont
  {Ludlow}}\ and\ \bibinfo {author} {\bibfnamefont {R.~E.}\ \bibnamefont
  {Angulo}},\ }\href {\doibase 10.1093/mnrasl/slw216} {\bibfield  {journal}
  {\bibinfo  {journal} {Mon. Not. Roy. Astron. Soc.}\ }\textbf {\bibinfo
  {volume} {465}},\ \bibinfo {pages} {L84} (\bibinfo {year} {2017})},\ \Eprint
  {http://arxiv.org/abs/1610.04620} {arXiv:1610.04620 [astro-ph.CO]}
  \BibitemShut {NoStop}%
\bibitem [{\citenamefont {Schmidt}(2016)}]{Schmidt:2015gwz}%
  \BibitemOpen
  \bibfield  {author} {\bibinfo {author} {\bibfnamefont {F.}~\bibnamefont
  {Schmidt}},\ }\href {\doibase 10.1103/PhysRevD.93.063512} {\bibfield
  {journal} {\bibinfo  {journal} {Phys. Rev.}\ }\textbf {\bibinfo {volume}
  {D93}},\ \bibinfo {pages} {063512} (\bibinfo {year} {2016})},\ \Eprint
  {http://arxiv.org/abs/1511.02231} {arXiv:1511.02231 [astro-ph.CO]}
  \BibitemShut {NoStop}%
\bibitem [{\citenamefont {Seljak}\ and\ \citenamefont
  {Vlah}(2015)}]{Seljak:2015rea}%
  \BibitemOpen
  \bibfield  {author} {\bibinfo {author} {\bibfnamefont {U.}~\bibnamefont
  {Seljak}}\ and\ \bibinfo {author} {\bibfnamefont {Z.}~\bibnamefont {Vlah}},\
  }\href {\doibase 10.1103/PhysRevD.91.123516} {\bibfield  {journal} {\bibinfo
  {journal} {Phys. Rev.}\ }\textbf {\bibinfo {volume} {D91}},\ \bibinfo {pages}
  {123516} (\bibinfo {year} {2015})},\ \Eprint
  {http://arxiv.org/abs/1501.07512} {arXiv:1501.07512 [astro-ph.CO]}
  \BibitemShut {NoStop}%
\bibitem [{\citenamefont {Hand}\ \emph {et~al.}(2017)\citenamefont {Hand},
  \citenamefont {Seljak}, \citenamefont {Beutler},\ and\ \citenamefont
  {Vlah}}]{Hand:2017ilm}%
  \BibitemOpen
  \bibfield  {author} {\bibinfo {author} {\bibfnamefont {N.}~\bibnamefont
  {Hand}}, \bibinfo {author} {\bibfnamefont {U.}~\bibnamefont {Seljak}},
  \bibinfo {author} {\bibfnamefont {F.}~\bibnamefont {Beutler}}, \ and\
  \bibinfo {author} {\bibfnamefont {Z.}~\bibnamefont {Vlah}},\ }\href {\doibase
  10.1088/1475-7516/2017/10/009} {\bibfield  {journal} {\bibinfo  {journal}
  {JCAP}\ }\textbf {\bibinfo {volume} {1710}},\ \bibinfo {pages} {009}
  (\bibinfo {year} {2017})},\ \Eprint {http://arxiv.org/abs/1706.02362}
  {arXiv:1706.02362 [astro-ph.CO]} \BibitemShut {NoStop}%
\bibitem [{\citenamefont {Philcox}\ \emph {et~al.}(2020)\citenamefont
  {Philcox}, \citenamefont {Spergel},\ and\ \citenamefont
  {Villaescusa-Navarro}}]{Philcox:2020rpe}%
  \BibitemOpen
  \bibfield  {author} {\bibinfo {author} {\bibfnamefont {O.~H.~E.}\
  \bibnamefont {Philcox}}, \bibinfo {author} {\bibfnamefont {D.~N.}\
  \bibnamefont {Spergel}}, \ and\ \bibinfo {author} {\bibfnamefont
  {F.}~\bibnamefont {Villaescusa-Navarro}},\ }\href {\doibase
  10.1103/PhysRevD.101.123520} {\bibfield  {journal} {\bibinfo  {journal}
  {Phys. Rev. D}\ }\textbf {\bibinfo {volume} {101}},\ \bibinfo {pages}
  {123520} (\bibinfo {year} {2020})},\ \Eprint
  {http://arxiv.org/abs/2004.09515} {arXiv:2004.09515 [astro-ph.CO]}
  \BibitemShut {NoStop}%
\bibitem [{\citenamefont {Cooray}\ and\ \citenamefont
  {Sheth}(2002)}]{Cooray:2002dia}%
  \BibitemOpen
  \bibfield  {author} {\bibinfo {author} {\bibfnamefont {A.}~\bibnamefont
  {Cooray}}\ and\ \bibinfo {author} {\bibfnamefont {R.~K.}\ \bibnamefont
  {Sheth}},\ }\href {\doibase 10.1016/S0370-1573(02)00276-4} {\bibfield
  {journal} {\bibinfo  {journal} {Phys. Rept.}\ }\textbf {\bibinfo {volume}
  {372}},\ \bibinfo {pages} {1} (\bibinfo {year} {2002})},\ \Eprint
  {http://arxiv.org/abs/astro-ph/0206508} {arXiv:astro-ph/0206508 [astro-ph]}
  \BibitemShut {NoStop}%
\bibitem [{\citenamefont {{Ginzburg}}\ \emph {et~al.}(2017)\citenamefont
  {{Ginzburg}}, \citenamefont {{Desjacques}},\ and\ \citenamefont
  {{Chan}}}]{2017PhRvD..96h3528G}%
  \BibitemOpen
  \bibfield  {author} {\bibinfo {author} {\bibfnamefont {D.}~\bibnamefont
  {{Ginzburg}}}, \bibinfo {author} {\bibfnamefont {V.}~\bibnamefont
  {{Desjacques}}}, \ and\ \bibinfo {author} {\bibfnamefont {K.~C.}\
  \bibnamefont {{Chan}}},\ }\href {\doibase 10.1103/PhysRevD.96.083528}
  {\bibfield  {journal} {\bibinfo  {journal} {\prd}\ }\textbf {\bibinfo
  {volume} {96}},\ \bibinfo {eid} {083528} (\bibinfo {year} {2017})},\ \Eprint
  {http://arxiv.org/abs/1706.08738} {arXiv:1706.08738 [astro-ph.CO]}
  \BibitemShut {NoStop}%
\bibitem [{\citenamefont {Umeh}\ \emph {et~al.}(2021)\citenamefont {Umeh},
  \citenamefont {Maartens}, \citenamefont {Padmanabhan},\ and\ \citenamefont
  {Camera}}]{Umeh:2021xqm}%
  \BibitemOpen
  \bibfield  {author} {\bibinfo {author} {\bibfnamefont {O.}~\bibnamefont
  {Umeh}}, \bibinfo {author} {\bibfnamefont {R.}~\bibnamefont {Maartens}},
  \bibinfo {author} {\bibfnamefont {H.}~\bibnamefont {Padmanabhan}}, \ and\
  \bibinfo {author} {\bibfnamefont {S.}~\bibnamefont {Camera}},\ }\href
  {\doibase 10.1088/1475-7516/2021/06/027} {\bibfield  {journal} {\bibinfo
  {journal} {JCAP}\ }\textbf {\bibinfo {volume} {06}},\ \bibinfo {pages} {027}
  (\bibinfo {year} {2021})},\ \Eprint {http://arxiv.org/abs/2102.06116}
  {arXiv:2102.06116 [astro-ph.CO]} \BibitemShut {NoStop}%
\bibitem [{\citenamefont {Takahashi}\ \emph {et~al.}(2012)\citenamefont
  {Takahashi}, \citenamefont {Sato}, \citenamefont {Nishimichi}, \citenamefont
  {Taruya},\ and\ \citenamefont {Oguri}}]{Takahashi:2012em}%
  \BibitemOpen
  \bibfield  {author} {\bibinfo {author} {\bibfnamefont {R.}~\bibnamefont
  {Takahashi}}, \bibinfo {author} {\bibfnamefont {M.}~\bibnamefont {Sato}},
  \bibinfo {author} {\bibfnamefont {T.}~\bibnamefont {Nishimichi}}, \bibinfo
  {author} {\bibfnamefont {A.}~\bibnamefont {Taruya}}, \ and\ \bibinfo {author}
  {\bibfnamefont {M.}~\bibnamefont {Oguri}},\ }\href {\doibase
  10.1088/0004-637X/761/2/152} {\bibfield  {journal} {\bibinfo  {journal}
  {Astrophys. J.}\ }\textbf {\bibinfo {volume} {761}},\ \bibinfo {pages} {152}
  (\bibinfo {year} {2012})},\ \Eprint {http://arxiv.org/abs/1208.2701}
  {arXiv:1208.2701 [astro-ph.CO]} \BibitemShut {NoStop}%
\bibitem [{\citenamefont {Mead}\ and\ \citenamefont
  {Verde}(2021)}]{Mead:2020qdk}%
  \BibitemOpen
  \bibfield  {author} {\bibinfo {author} {\bibfnamefont {A.~J.}\ \bibnamefont
  {Mead}}\ and\ \bibinfo {author} {\bibfnamefont {L.}~\bibnamefont {Verde}},\
  }\href {\doibase 10.1093/mnras/stab748} {\bibfield  {journal} {\bibinfo
  {journal} {Mon. Not. Roy. Astron. Soc.}\ }\textbf {\bibinfo {volume} {503}},\
  \bibinfo {pages} {3095} (\bibinfo {year} {2021})},\ \Eprint
  {http://arxiv.org/abs/2011.08858} {arXiv:2011.08858 [astro-ph.CO]}
  \BibitemShut {NoStop}%
\bibitem [{\citenamefont {Nishimichi}\ \emph {et~al.}(2019)\citenamefont
  {Nishimichi} \emph {et~al.}}]{Nishimichi:2018etk}%
  \BibitemOpen
  \bibfield  {author} {\bibinfo {author} {\bibfnamefont {T.}~\bibnamefont
  {Nishimichi}} \emph {et~al.},\ }\href {\doibase 10.3847/1538-4357/ab3719}
  {\bibfield  {journal} {\bibinfo  {journal} {Astrophys. J.}\ }\textbf
  {\bibinfo {volume} {884}},\ \bibinfo {pages} {29} (\bibinfo {year} {2019})},\
  \Eprint {http://arxiv.org/abs/1811.09504} {arXiv:1811.09504 [astro-ph.CO]}
  \BibitemShut {NoStop}%
\bibitem [{\citenamefont {Ade}\ \emph {et~al.}(2016)\citenamefont {Ade} \emph
  {et~al.}}]{Ade:2015zua}%
  \BibitemOpen
  \bibfield  {author} {\bibinfo {author} {\bibfnamefont {P.~A.~R.}\
  \bibnamefont {Ade}} \emph {et~al.} (\bibinfo {collaboration} {Planck}),\
  }\href {\doibase 10.1051/0004-6361/201525941} {\bibfield  {journal} {\bibinfo
   {journal} {Astron. Astrophys.}\ }\textbf {\bibinfo {volume} {594}},\
  \bibinfo {pages} {A15} (\bibinfo {year} {2016})},\ \Eprint
  {http://arxiv.org/abs/1502.01591} {arXiv:1502.01591 [astro-ph.CO]}
  \BibitemShut {NoStop}%
\bibitem [{\citenamefont {More}\ \emph {et~al.}(2015)\citenamefont {More},
  \citenamefont {Diemer},\ and\ \citenamefont {Kravtsov}}]{More:2015ufa}%
  \BibitemOpen
  \bibfield  {author} {\bibinfo {author} {\bibfnamefont {S.}~\bibnamefont
  {More}}, \bibinfo {author} {\bibfnamefont {B.}~\bibnamefont {Diemer}}, \ and\
  \bibinfo {author} {\bibfnamefont {A.}~\bibnamefont {Kravtsov}},\ }\href
  {\doibase 10.1088/0004-637X/810/1/36} {\bibfield  {journal} {\bibinfo
  {journal} {Astrophys. J.}\ }\textbf {\bibinfo {volume} {810}},\ \bibinfo
  {pages} {36} (\bibinfo {year} {2015})},\ \Eprint
  {http://arxiv.org/abs/1504.05591} {arXiv:1504.05591 [astro-ph.CO]}
  \BibitemShut {NoStop}%
\bibitem [{\citenamefont {Asgari}\ \emph {et~al.}(2023)\citenamefont {Asgari},
  \citenamefont {Mead},\ and\ \citenamefont {Heymans}}]{Asgari:2023mej}%
  \BibitemOpen
  \bibfield  {author} {\bibinfo {author} {\bibfnamefont {M.}~\bibnamefont
  {Asgari}}, \bibinfo {author} {\bibfnamefont {A.~J.}\ \bibnamefont {Mead}}, \
  and\ \bibinfo {author} {\bibfnamefont {C.}~\bibnamefont {Heymans}},\
  }\href@noop {} {\  (\bibinfo {year} {2023})},\ \Eprint
  {http://arxiv.org/abs/2303.08752} {arXiv:2303.08752 [astro-ph.CO]}
  \BibitemShut {NoStop}%
\bibitem [{\citenamefont {Dalal}\ \emph {et~al.}(2008)\citenamefont {Dalal},
  \citenamefont {White}, \citenamefont {Bond},\ and\ \citenamefont
  {Shirokov}}]{Dalal:2008zd}%
  \BibitemOpen
  \bibfield  {author} {\bibinfo {author} {\bibfnamefont {N.}~\bibnamefont
  {Dalal}}, \bibinfo {author} {\bibfnamefont {M.}~\bibnamefont {White}},
  \bibinfo {author} {\bibfnamefont {J.}~\bibnamefont {Bond}}, \ and\ \bibinfo
  {author} {\bibfnamefont {A.}~\bibnamefont {Shirokov}},\ }\href {\doibase
  10.1086/591512} {\bibfield  {journal} {\bibinfo  {journal} {Astrophys. J.}\
  }\textbf {\bibinfo {volume} {687}},\ \bibinfo {pages} {12} (\bibinfo {year}
  {2008})},\ \Eprint {http://arxiv.org/abs/0803.3453} {arXiv:0803.3453
  [astro-ph]} \BibitemShut {NoStop}%
\bibitem [{\citenamefont {Tinker}\ \emph {et~al.}(2008)\citenamefont {Tinker},
  \citenamefont {Kravtsov}, \citenamefont {Klypin}, \citenamefont {Abazajian},
  \citenamefont {Warren}, \citenamefont {Yepes}, \citenamefont {Gottlober},\
  and\ \citenamefont {Holz}}]{Tinker:2008ff}%
  \BibitemOpen
  \bibfield  {author} {\bibinfo {author} {\bibfnamefont {J.~L.}\ \bibnamefont
  {Tinker}}, \bibinfo {author} {\bibfnamefont {A.~V.}\ \bibnamefont
  {Kravtsov}}, \bibinfo {author} {\bibfnamefont {A.}~\bibnamefont {Klypin}},
  \bibinfo {author} {\bibfnamefont {K.}~\bibnamefont {Abazajian}}, \bibinfo
  {author} {\bibfnamefont {M.~S.}\ \bibnamefont {Warren}}, \bibinfo {author}
  {\bibfnamefont {G.}~\bibnamefont {Yepes}}, \bibinfo {author} {\bibfnamefont
  {S.}~\bibnamefont {Gottlober}}, \ and\ \bibinfo {author} {\bibfnamefont
  {D.~E.}\ \bibnamefont {Holz}},\ }\href {\doibase 10.1086/591439} {\bibfield
  {journal} {\bibinfo  {journal} {Astrophys. J.}\ }\textbf {\bibinfo {volume}
  {688}},\ \bibinfo {pages} {709} (\bibinfo {year} {2008})},\ \Eprint
  {http://arxiv.org/abs/0803.2706} {arXiv:0803.2706 [astro-ph]} \BibitemShut
  {NoStop}%
\bibitem [{\citenamefont {Dai}\ \emph {et~al.}(2022)\citenamefont {Dai},
  \citenamefont {Starkman},\ and\ \citenamefont {Stojkovic}}]{Dai:2022had}%
  \BibitemOpen
  \bibfield  {author} {\bibinfo {author} {\bibfnamefont {D.-C.}\ \bibnamefont
  {Dai}}, \bibinfo {author} {\bibfnamefont {G.}~\bibnamefont {Starkman}}, \
  and\ \bibinfo {author} {\bibfnamefont {D.}~\bibnamefont {Stojkovic}},\ }\href
  {\doibase 10.1103/PhysRevD.105.104067} {\bibfield  {journal} {\bibinfo
  {journal} {Phys. Rev. D}\ }\textbf {\bibinfo {volume} {105}},\ \bibinfo
  {pages} {104067} (\bibinfo {year} {2022})},\ \Eprint
  {http://arxiv.org/abs/2201.06034} {arXiv:2201.06034 [astro-ph.GA]}
  \BibitemShut {NoStop}%
\bibitem [{\citenamefont {Lewis}\ \emph {et~al.}(2000)\citenamefont {Lewis},
  \citenamefont {Challinor},\ and\ \citenamefont {Lasenby}}]{Lewis:1999bs}%
  \BibitemOpen
  \bibfield  {author} {\bibinfo {author} {\bibfnamefont {A.}~\bibnamefont
  {Lewis}}, \bibinfo {author} {\bibfnamefont {A.}~\bibnamefont {Challinor}}, \
  and\ \bibinfo {author} {\bibfnamefont {A.}~\bibnamefont {Lasenby}},\ }\href
  {\doibase 10.1086/309179} {\bibfield  {journal} {\bibinfo  {journal} {\apj}\
  }\textbf {\bibinfo {volume} {538}},\ \bibinfo {pages} {473} (\bibinfo {year}
  {2000})},\ \Eprint {http://arxiv.org/abs/astro-ph/9911177}
  {arXiv:astro-ph/9911177 [astro-ph]} \BibitemShut {NoStop}%
\bibitem [{\citenamefont {Peacock}\ and\ \citenamefont
  {Smith}(2000)}]{Peacock:2000qk}%
  \BibitemOpen
  \bibfield  {author} {\bibinfo {author} {\bibfnamefont {J.~A.}\ \bibnamefont
  {Peacock}}\ and\ \bibinfo {author} {\bibfnamefont {R.~E.}\ \bibnamefont
  {Smith}},\ }\href {\doibase 10.1046/j.1365-8711.2000.03779.x} {\bibfield
  {journal} {\bibinfo  {journal} {Mon. Not. Roy. Astron. Soc.}\ }\textbf
  {\bibinfo {volume} {318}},\ \bibinfo {pages} {1144} (\bibinfo {year}
  {2000})},\ \Eprint {http://arxiv.org/abs/astro-ph/0005010}
  {arXiv:astro-ph/0005010 [astro-ph]} \BibitemShut {NoStop}%
\bibitem [{\citenamefont {Abbott}\ \emph {et~al.}(2022)\citenamefont {Abbott}
  \emph {et~al.}}]{DES:2021wwk}%
  \BibitemOpen
  \bibfield  {author} {\bibinfo {author} {\bibfnamefont {T.~M.~C.}\
  \bibnamefont {Abbott}} \emph {et~al.} (\bibinfo {collaboration} {DES}),\
  }\href {\doibase 10.1103/PhysRevD.105.023520} {\bibfield  {journal} {\bibinfo
   {journal} {Phys. Rev. D}\ }\textbf {\bibinfo {volume} {105}},\ \bibinfo
  {pages} {023520} (\bibinfo {year} {2022})},\ \Eprint
  {http://arxiv.org/abs/2105.13549} {arXiv:2105.13549 [astro-ph.CO]}
  \BibitemShut {NoStop}%
\bibitem [{\citenamefont {Heymans}\ \emph {et~al.}(2021)\citenamefont {Heymans}
  \emph {et~al.}}]{Heymans:2020gsg}%
  \BibitemOpen
  \bibfield  {author} {\bibinfo {author} {\bibfnamefont {C.}~\bibnamefont
  {Heymans}} \emph {et~al.},\ }\href {\doibase 10.1051/0004-6361/202039063}
  {\bibfield  {journal} {\bibinfo  {journal} {Astron. Astrophys.}\ }\textbf
  {\bibinfo {volume} {646}},\ \bibinfo {pages} {A140} (\bibinfo {year}
  {2021})},\ \Eprint {http://arxiv.org/abs/2007.15632} {arXiv:2007.15632
  [astro-ph.CO]} \BibitemShut {NoStop}%
\bibitem [{\citenamefont {LoVerde}\ and\ \citenamefont
  {Afshordi}(2008)}]{LoVerde:2008re}%
  \BibitemOpen
  \bibfield  {author} {\bibinfo {author} {\bibfnamefont {M.}~\bibnamefont
  {LoVerde}}\ and\ \bibinfo {author} {\bibfnamefont {N.}~\bibnamefont
  {Afshordi}},\ }\href {\doibase 10.1103/PhysRevD.78.123506} {\bibfield
  {journal} {\bibinfo  {journal} {Phys. Rev.}\ }\textbf {\bibinfo {volume}
  {D78}},\ \bibinfo {pages} {123506} (\bibinfo {year} {2008})},\ \Eprint
  {http://arxiv.org/abs/0809.5112} {arXiv:0809.5112 [astro-ph]} \BibitemShut
  {NoStop}%
\end{thebibliography}%

\end{document}